\documentclass[final]{siamonline171218}

\usepackage{multirow}
\usepackage{tabulary}
\usepackage{colortbl}

\usepackage{lipsum}
\usepackage{amsfonts}
\usepackage{graphicx}
\usepackage{epstopdf}
\usepackage{amssymb}
\usepackage{amsmath}
\usepackage{algorithm}
\usepackage[noend]{algpseudocode}
\usepackage{diagbox}

\graphicspath{{Figures/}}

\usepackage[utf8]{inputenc}
\usepackage{multirow}
\usepackage{hhline}
\usepackage{xcolor}
\usepackage{epstopdf}
\usepackage{caption}
\usepackage{subcaption}
\usepackage{bm}
\usepackage{pgfplots}
\usepackage{footnote}
\usepackage{stfloats}
\usepackage{placeins}
\usepackage{color,soul}
\usepackage{dsfont}
\usepackage{pbox}
\usepackage{enumerate}
\usepackage{etextools}
\usepackage{tabulary}
\usepackage{dsfont}
\usepackage{mathtools}
\usepackage{enumitem}
\usepackage{parskip}
\usepackage[algo2e,algoruled,boxed,lined]{algorithm2e}

\def\0{{\mathbf 0}}
\def\1{{\mathbf 1}}
\def\x{{\mathbf x}}

\def\T{{\mathbf T}}
\def\X{{\mathbf X}}
\def\v{{\mathbf v}}

\def\u{ \mathbf{u}}

\def\y{{\mathbf y}}

\def\z{{\mathbf z}}

\def\I{{\mathbf I}}
\def\u{{\mathbf u}}
\def\U{{\mathbf U}}

\def\D{{\mathbf D}}

\def\W{{\mathbf W}}

\def\H{{\mathbf{H}}}

\def\L{{\mathbf L}}

\def\R{{\mathbf R}}

\def\V{{\mathbf V}}

\DeclareMathOperator*{\argmin}{arg\,min}
\DeclareMathOperator*{\const}{const}
\DeclareMathOperator*{\dist}{dist}
\DeclareMathOperator*{\subjectto}{s.t.}

\setlist[enumerate]{parsep=\smallskipamount}

\makeatletter
\def\BState{\State\hskip-\ALG@thistlm}
\makeatother

\headers{Image Denoising: The Deep Learning Revolution and Beyond }{M. Elad, B. Kawar and G. Vaksman}

\title{Image Denoising: The Deep Learning Revolution and Beyond \\ -- A Survey Paper -- }

\author{Michael~Elad, Bahjat~Kawar and Gregory~Vaksman \\ The Computer Science Department, Technion -- Israel Institute of Technology \\ email: \{elad,bahjat.kawar,grishavak\}@cs.technion.ac.il}

\begin{document}
\maketitle


\begin{abstract}

Image denoising -- removal of additive white Gaussian noise from an image -- is one of the oldest and most studied problems in image processing. An extensive work over several decades has led to thousands of papers on this subject, and to many well-performing algorithms for this task. Indeed, ten years ago, these achievements have led some researchers to suspect that “Denoising is Dead”, in the sense that all that can be achieved in this domain has already been obtained. However, this turned out to be far from the truth, with the penetration of deep learning (DL) into the realm of image processing. The era of DL brought a revolution to image denoising, both by taking the lead in today’s ability for noise suppression in images, and by broadening the scope of denoising problems being treated. Our paper starts by describing this evolution,  highlighting in particular the tension and synergy that exist between classical approaches and modern Artificial Intelligence (AI) alternatives in design of image denoisers.

The recent transitions in the field of image denoising go far beyond the ability to design better denoisers. In the second part of this paper we focus on recently discovered abilities and prospects of image denoisers. We expose the possibility of using image denoisers for service of other problems, such as regularizing general inverse problems and serving as the prime engine in diffusion-based image synthesis. We also unveil the (strange?) idea that denoising and other inverse problems might not have a unique solution, as common algorithms would have us believe. Instead, we describe constructive ways to produce randomized and diverse high perceptual quality results for inverse problems, all fueled by the progress that DL brought to image denoising.

This is a survey paper, and its prime goal is to provide a broad view of the history of the field of image denoising and closely related topics in image processing. Our aim is to give a better context to recent discoveries, and to the influence of the AI revolution in our domain.   

\end{abstract}




\section{Introduction}

Within the wide fields of image processing and computational imaging, the task of image denoising has been given an exceptionally large attention over the past several decades. Indeed, noise suppression in images is one of the oldest and most studied problems in these fields, with numerous papers offering diverse algorithms, analysis of this task in various forms, or extensions of it.\footnote{See Figure \ref{fig:Papers} for the quantities of denoising related papers over the years.} A substantial portion of the proposed denoising techniques has been dedicated to the removal of Additive White Gaussian Noise (AWGN) from images, while there are other contributions that target different noise distributions, \textit{e.g.} Poisson, salt-and-pepper, and more.  

Removal of noise from an image is an actual necessity that comes up with practically every imaging sensor~\cite{martinec2008noise}. However, the interest in this problem goes far beyond this practical need -- image denoising is the simplest inverse problem, and as such, it has been recognized over the years as the perfect test-bed for assessing new ideas that are often brought to image processing. In recent years this appeal has further widened with the realization that denoisers can serve other imaging needs~\cite{PnP,RED,song2019generative}. 

The years 1980 -- 2010 have seen consistently improving denoising algorithms, many of which relying on the Bayesian point of view. This progress has been geared by an evolution of image priors that form the backbone of the overall progress in image processing. This path, which we will refer to as the \emph{classical era}, started with the early $L_2$-based regularization, proceeding to robust statistics, moving to the introduction of wavelets, and the later deployment of partial differential equations to imaging tasks, and this continued all the way to sparse modeling, patch-based methods, and low-rank structure assumptions\footnote{As referencing this is too long, we provide specific citations to each of these in later sections.}. This extensive work over several decades has led to many well-performing denoising algorithms, and to a compelling and rich scientific field. In fact, ten years ago, these glorious achievements have led some researchers to consider the possibility that ``Denoising is Dead'', believing that the existing solutions are already touching the achievable performance ceiling~\cite{PeymanDiD,levin2011natural,levin2012patch}. 

The past decade has brought a paradigm shift to the general topic of data processing due to the emergence of the Artificial Intelligence (AI) revolution. The great success of this deep learning (DL) trend has also introduced a reformation to the broad field of image processing, and to image denoising in particular. These new winds led to novel techniques for designing better performing denoisers~\cite{chen2016trainable, DnCNN, liu2018non, zhang2018ffdnet, tai2017memnet, UDNet, anwar2019real, zhang2020residual, liang2021swinir}, and discovering new and more daring ways for deploying them and broadening their scope~\cite{abdelhamed2018high, liu2021invertible, zamir2021multi, vaksman2022patch, godard2018deep, liang2020decoupled, kawar2021stochastic, ohayon2021high, guo2021joint}.
These days, deep-learning based denoisers are at the very top in their ability for noise suppression in images (see \textit{e.g.} \cite{DnCNN,liang2021swinir,zamir2022restormer}, leaving no competitive room for the classical alternatives). 

In parallel to the above and seemingly detached from the deep learning activity, image denoising has been also a topic of investigation and discoveries of a different flavor: Harnessing denoiser engines for other imaging tasks. This started with the surprising idea that a good performing denoiser can serve as a prior, offering a highly effective regularization to inverse problems~\cite{PnP,RED,brifman2016turning,kamilov2017plug,tirer2018image,sun2019online,mataev2019deepred,teodoro2019image,chen2021deep,cohen2021has}. This continued with the discovery that such denoisers can also be used for randomly synthesizing images by offering a practical sampling from the prior distribution of images, this way posing a potent competition to Generative Adversarial Networks (GANs) and other image generation methods~\cite{song2019generative, song2020improved, song2021sde, ho2020denoising, vahdat2021score, guided_diffusion, ho2021classifier, kawar2022enhancing, ho2022cascaded}. 

An intriguing sequel to the above synthesis revelation is the idea that solution of inverse problems could be revisited and posed as a sampling task from the posterior distribution of the image given the measurements, thus resorting again to image denoisers as the means for obtaining these solutions. This very recent line of work unveiled the daring idea that denoising and other inverse problems might not have a unique solution, as common algorithms would have us believe~\cite{ohayon2021high, kawar2021stochastic, kadkhodaie2021stochastic, kawar2021snips, ohayon2022reasons}. Instead, this sampling approach has been shown to lead to constructive ways for producing randomized and diverse \emph{high perceptual quality} results for inverse problems, exposing as a byproduct the inner uncertainty in such tasks. 

All the above achievements have been strongly influenced and fueled by the progress that DL brought to image denoising. Adopting a wider perspective, image denoising these days has new horizons, and if any conclusion can be drawn from these recent accomplishments, it would be that this field is a very much alive playground with great challenges and prospects. This paper aims to disclose and detail the compelling story drawn above. Our prime goal is to provide a broad view of the history of the field of image denoising and closely related topics in image processing, give a better context to recent discoveries, and  highlight the influence of the AI revolution in our domain. 

We start our journey in Section \ref{sec:Back} by clearly defining the image denoising task, discussing its ill-posed nature, and demonstrating its appeal over the years. We proceed in Sections \ref{sec:Classic} and \ref{sec:DL} by describing the evolution of image denoisers, from the classical era to the deep-learning-based alternatives. Section \ref{sec:Synergy} highlights the tension and the possible synergy that exists between classical approaches and modern Artificial Intelligence (AI) alternatives in design of image denoisers. 

In the second part of the paper we change gears and move to discuss three recent discoveries that consider image denoisers as building blocks for other needs. We start broadly in Section \ref{sec:Discoveries} by defining the denoiser engine and its properties, and set the stage for the presentation of these three discoveries. We proceed in Section \ref{sec:IP} by discussing the ability to deploy these engines for regularizing inverse problems. Section \ref{sec:Synthesis} exposes the possibility of synthesizing images using such denoisers, and Section \ref{sec:HPC} presents the notion of targeting perfect perceptual quality outcomes in image denoising and inverse problems by sampling from the posterior distribution. We conclude this paper in Section \ref{sec:conclusion} with an attempt to point to open questions and potential research directions.

\emph{Disclaimer:} While this paper aims to present a survey on the various ups and downs that the field of image denoising has gone through over the years, it would be simply impossible to do justice to all the published literature in this domain. We apologize if some papers are omitted from our references, as we attempt to mark the critical milestones in the history of this field. The interested reader is referred to~\cite{lebrun2012secrets,milanfar2012tour,jain2016survey,bertalmio2018denoising,fan2019brief,tian2020deep} for other surveys with different orientations.


\section{Image Denoising -- Background}
\label{sec:Back}

\subsection{Problem Definition}
\label{sec:ProbDef}
Our story must start with a proper definition of the denoising problem, and this will also serve the need for defining our notations hereafter. 
An ideal image\footnote{For simplicity of the discussion, assume that we refer to grayscale images. Addressing color is discussed shortly in Section \ref{sec:DL}.} $\x \in \mathbb{R}^N$ is assumed to be drawn from the image manifold, represented by the probability density function $p(\x)$. Our measurement is the vector $\y\in \mathbb{R}^N$, given by 
\begin{eqnarray}
\label{eq:denoising}
\y=\x+\v,
\end{eqnarray}
where $\v \in \mathbb{R}^N$ is a zero-mean independent and identically distributed (i.i.d.) Gaussian noise, \textit{i.e.} $\v \sim {\cal N}(\0,\sigma ^2 \I)$. The denoising task is the recovery of $\x$ from $\y$ with the knowledge of $\sigma$, and a denoiser is thus a function of the form $\hat{\x}=D(\y,\sigma)$. 

While there are many ways for assessing the performance of such denoisers, the most common one is the Mean-Squared-Error (MSE) measure, 
\begin{eqnarray}
\label{eq:MSE}
MSE = \mathbb{E}\left( \| \x-{\hat \x} \|_2^2\right)=\mathbb{E}\left( \| \x- D(\y,\sigma) \|_2^2\right),
\end{eqnarray}
where the expectation is taken over the image distribution. A well-known result in estimation theory states that the best denoising with respect to this measure (\textit{i.e.}, achieving the Minimum MSE, thus referred to as MMSE) is given by~\cite{MMSE}, 
\begin{eqnarray}
\label{eq:MMSE}
{\hat \x}_{MMSE} = \mathbb{E}\left( \x | \y \right).
\end{eqnarray}
This formula is misleadingly simple in its concise form, as designing a denoiser that achieves MMSE is quite challenging and oftentimes simply impossible. By the way, the curious reader may wonder why are we emphasizing the MSE measure and the MMSE denoiser. The answer will be carefully unfolded in the later parts of the paper, where these choices play a critical role. A brief note about this appears later in this section.  

How hard is it to denoise an image? How complicated could it be? Again, the simplicity of the problem definition is illusive, as this task is highly tough and in fact ill-posed. One could easily design a filtering method for attenuating and suppressing the noise in $\y$, but such a process is very likely to ruin the image content as well, losing small details, sacrificing edges, damaging fine textures, and more. 

\subsection{The Gaussianity Assumption}

In the problem definition above we focused on a very specific case of a zero-mean i.i.d. Gaussian noise contamination. The natural question arising is why are we restricting the discussion to this case? A brief inspection of the literature on image denoising reveals that this noise model is very popular, covered by most of the developed algorithms. Where this popularity comes from? Several answers come to mind:
\begin{itemize}
    \item {\bf Central Limit Theorem:} Noise in imaging may arise due to many physical reasons, and their accumulation leads often to a Gaussian distribution of the form discussed above, as an empirical manifestation of the Central Limit Theorem~\cite{chebyshev, Ibragimov1971IndependentAS}. As such, rather than modelling the intricate noise origins, a Gaussian assumption offers a blessed simplification for the later analysis and algorithm development in this field.
    
    \item {\bf The Poisson Alternative:} One might rightfully argue that the proper distribution to address for imaging noise would be the Poisson one, as imaging sensors essentially count photons, and their arrival is of Poissonian nature~\cite{PoissonNoise}. While this argument is indeed correct, when photon counts are high, the Poisson distribution becomes a Gaussian one~\cite{HighPoissonCount}. If the counts are low, a variance stabilizing transform, such as \emph{Anscombe}~\cite{anscombe1948transformation}, can turn these measurements into additive Gaussian contaminated, again resorting to the Gaussianity regime~\cite{dupe2009proximal,zhang2008wavelets,rond2016poisson,makitalo2010optimal,azzari2016variance,talbot2009efficient,zhang2019vst}.
    
    \item {\bf Mathematical Elegance:} The Gaussian case is easily modeled, and consequent formulations become simple and elegant. Such is the case with the log-likelihood function $p(\y|\x)$ and other related derivations that will be shown in subsequent sections.
    
    \item {\bf MMSE Denoiser Engines:} Our last argument for the Gaussianity assumption is quite surprising and unfamiliar to many in our field. As it turns out, an MMSE denoiser for the removal of zero-mean i.i.d. Gaussian noise is of great theoretical importance. Such an engine has critical properties that enable its deployment as a prior (see Section \ref{sec:IP}) for inverse problems. In addition, and perhaps more importantly, such denoisers have strong theoretical ties to the \emph{score function}~\cite{song2019generative}, a fact that will be highlighted and exploited in Sections \ref{sec:Synthesis}-\ref{sec:HPC-IP}.

\end{itemize}


\subsection{Extensions of Image Denoising}

There are many variations to the core image denoising task mentioned above. These can be roughly divided into four sub-categories: (i) Handling different noise statistics; (ii) Addressing structured noise removal; (iii) Considering different and various visual content; and (iv) Posing different problem assumptions and settings. Lets us briefly describe each of these.

A natural extension of the original denoising problem posed above is to consider other noise statistics, such as Poisson (also referred to as shot-noise) denoising~\cite{figueiredo2010restoration,salmon2014poisson,giryes2014sparsity,deledalle2010poisson,zhang2008wavelets,vaksman2016patch,rond2016poisson,makitalo2010optimal,remez2018class,azzari2016variance}, salt-and-pepper noise removal~\cite{chan2005salt,Srinivasan2007ANF,Dong2007AND,Wang1999ProgressiveSM,Nikolova2004AVA}, treating mixtures of Poisson and Gaussian noise~\cite{luisier2010image,zhang2007multiscale,makitalo2010optimal}, and more. Other extensions consider structured noise, such as quantization noise in compression artifact removal~\cite{neville2006wavelet,luisier2010image, wang2013adaptive, dar2016postprocessing, gonzalez2018joint}, film-grain removal~\cite{Yan1997SignaldependentFG,yan1998film,dai2010film}, and textured or otherwise colored noise~\cite{goossens2009removal, matrecano2012improved, aggarwal2016hyperspectral, ponomarenko2008adaptive}. Another challenging task is noise removal in scenarios in which the noise is not spatially homogeneous, such as white noise with spatially varying $\sigma$~\cite{manjon2010adaptive, zhang2018ffdnet, zhou2020awgn, khowaja2021cascaded}. The inpainting problem~\cite{Elad2005SimultaneousCA,Mairal2008SparseRF,Xie2012ImageDA,Jin2015AnnihilatingFL} can be regarded as a special such case, where portions of the image are simply missing and need to be revived. These missing pixels can be regarded as contaminated by a very strong noise, while other regions of the image are reliably measured. 

The denoising task may assume a different setting altogether if the visual content is of different form. Such an example is noise reduction in bursts of snapshots~\cite{liu2014fast, godard2018deep, mildenhall2018burst, marinvc2019multi, ehret2019joint, liang2020decoupled, dudhane2022burst}, where several images are treated jointly. Somewhat similar yet different is the task of video denoising~\cite{maggioni2011video,arias2018video,arias2019kalman,tassano2019dvdnet,tassano2020fastdvdnet,Vaksman2021PatchCV,Song2022TempFormerTC,Maggioni2021EfficientMV,zamir2022restormer,Lei2020BlindVT}, in which we may seek online filtering of the incoming frames. When handling specific imaging types (\textit{e.g.}, microscopy~\cite{rajan2008improved, boulanger2009patch, bal2012dual, gokdaug2019image, zhang2019poisson, manifold2019denoising, lee2020mu, mannam2022real}, CT~\cite{li2014adaptive, chen2017low, yi2018sharpness, yang2018low, diwakar2020ct, yang2022low} and PET/SPECT imaging~\cite{christian2010dynamic, dutta2013non, gong2018pet, reymann2019u, zhou2020supervised, sun2022pix2pix}, and more), the algorithm design may require adequate adaptations to the data format (\textit{e.g.} treating 3D volumes~\cite{vatsa2009denoising, zhang2014denoising3d, zeng2018three, das2020adaptive, Luo_2021_ICCV}) or to the way it is captured. 

The last category of extensions has to do with our prior knowledge when addressing denoising tasks. Blind denoising~\cite{jain2008natural, liu2013single, lebrun2014noise, chen2018deep, mohan2019robust, soh2021deep, zhao2019simple, yue2019variational, guo2019toward} refers to the case in which the noise is known to be i.i.d. Gaussian, but $\sigma$ is unknown, and may be even spatially changing. A more complex situation is when the noise statistics are totally unknown~\cite{zhu2016noise, anwar2019real, abdelhamed2019ntire, xu2018trilateral}. In this context, a special case of great interest in recent years is removal of true noise from given images captured by digital cameras (\textit{e.g.}, cellphones)~\cite{wang2020practical, kim2014image, liu2021invertible, tran2020gan, wei2020physics, ignatov2021fast}. 


\subsection{The Interest in Image Denoising}

\begin{figure}[htbp]
    \centering
    \includegraphics[width=1\textwidth]{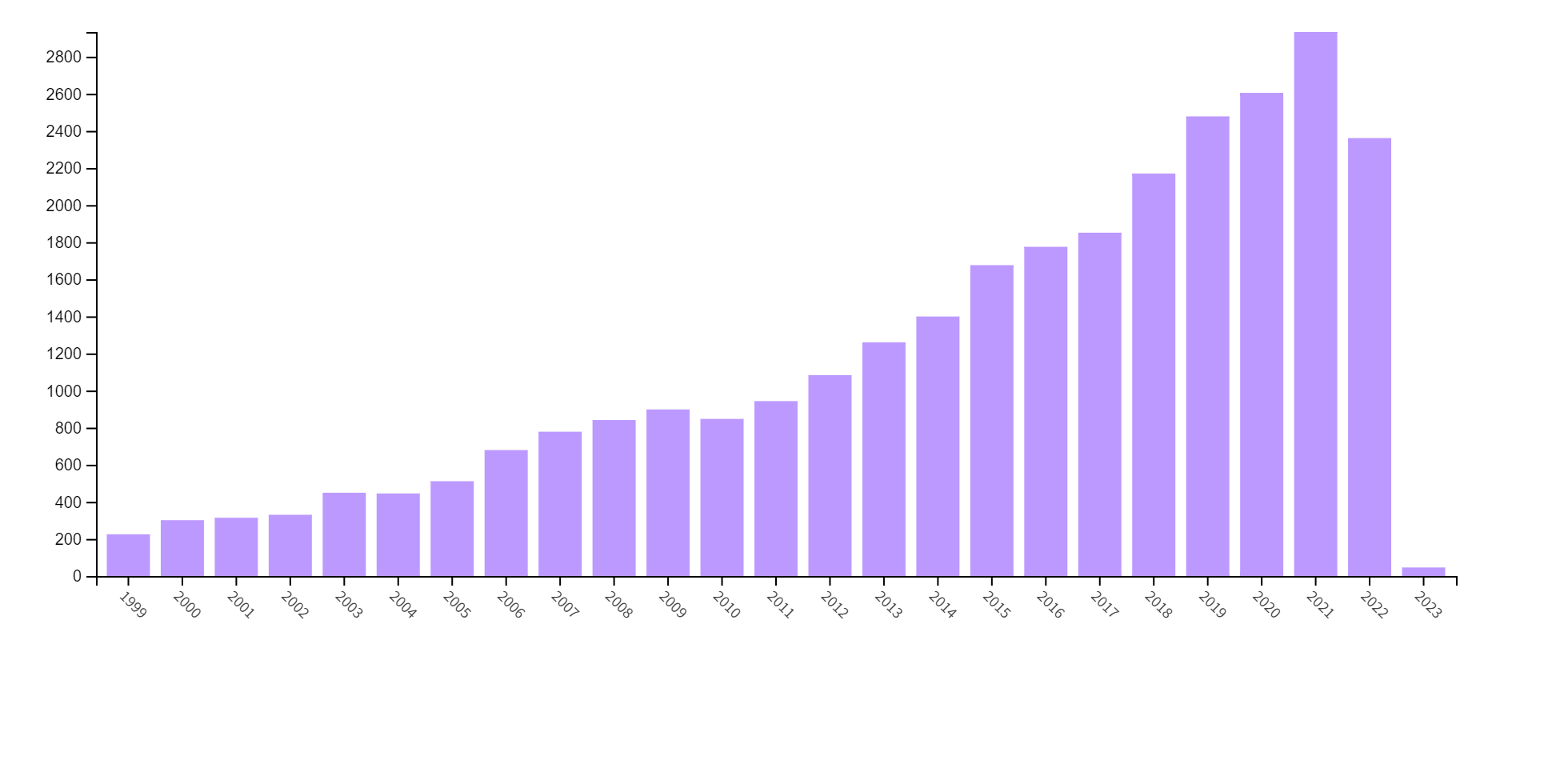}
    \caption{The number of papers on the image denoising topic over the years. This graph corresponds to the search \textsf{topic=((image or video ) and (denoising or (noise and remov) or clean))} performed on December 1st 2022 in Clarivate Web-of-Science (WoS). Note that the lower count in 2022 does not stand for a new trend, but rather caused by a delayed reporting of new papers.}
    \label{fig:Papers}
\end{figure}

Figure \ref{fig:Papers} presents a graph showing the number of papers that have been published each year on the topic of image denoising. Overall, nearly $30,000$ such papers are identified by Clarivate Web-Of-Science (WoS), published mostly in the past 25 years. As such, this is clearly one of the most heavily studied topics in image processing, and perhaps in exact sciences in general. Also evident from this graph is the consistent growth over the years in this topic. Where does this popularity come from? 

A prime reason to study image denoising is its practical relevance to imaging systems. Removal of noise from acquired visual information is an actual necessity that comes up with practically every imaging sensor~\cite{martinec2008noise}. Thus, various algorithms have been developed for implementation in image processing software packages and within the ISP (Image Signal Processor) -- the path that starts with the raw acquired data and ends with a high quality image -- of every digital camera~\cite{bosco2010signal, brooks2019unprocessing, wang2020practical, zhang2021rethinking, hu2021pseudo}. 

Beyond the obvious practical motivation described above, the interest in image denoising has other, more profound, roots. Image denoising is the simplest inverse problem, and as such, it has been recognized over the years as the perfect platform for assessing new ideas that are often brought to image processing. Indeed, all the major milestone advancements in image processing started with denoising experiments, so as to explore their validity to visual data. This was the case with Tikhonov-Arsenin's regularization theory~\cite{TikhonovBook,golub1999tikhonov}, Wavelets~\cite{mallat1999wavelet}, non--linear filtering based on partial differential equations~\cite{weickert1998pde,Guishard20022pde}, sparse modeling of data~\cite{bruckstein2009sparse,elad2010sparse}, and more. All these and many other sub-fields in imaging sciences saw image denoising as a critical first step in their diffusion into broad image processing tasks. We discuss these in greater details in the next section. 

The above two reasons for the popularity of image denoising may account for many of the published papers in this domain. However, the reason we have chosen to write this paper has to do with a third, and very different, origin of popularity. Image denoising has gained much interest and appeal in recent years due to the surprising realization that denoisers can serve other imaging needs, thus widening their scope and influence~\cite{PnP, RED, song2019generative, ho2020denoising, kawar2021snips}. This discovery relies on a fundamental theoretical connection between denoisers and the prior distribution of images~\cite{Miyasawa61, stein1981estimation, efron2011tweedie}. This bridge provides a solid and well-motivated approach to old and new tasks in image processing. In fact, this is the topic we shall be highlighting in the latter sections of our paper. We thus defer a more detailed explanation of these ideas. 


\section{Image Denoising -- The Classic Era}
\label{sec:Classic}

So far we have discussed image denoisers without concretely diving into the actual quest of their construction. So, how can we design an image denoiser? Not so surprisingly, the answer to this question has evolved and changed over the years, with the accumulated knowledge and the progress in signal and image processing. And still, we may broadly separate this progress in design of image denoisers into two eras - the classical one that started in the 70's and ended in the past decade, and the AI revolution era that started around 2012 and is very much vivid till this day. In this section we shall focus on the classical algorithms, and more specifically on the Bayesian point of view that played a key role in their creation. 


\subsection{The Bayesian Point of View for Design of Denoisers}
\label{sec:Bayes}

Starting with Equation (\ref{eq:denoising}), given the noisy image $\y$ and knowing that $\v \sim {\cal N}(\0,\sigma^2 \I)$, our goal is to estimate $\x$. A simple approach towards this task would be the Maximum-Likelihood Estimation (MLE)~\cite{murphy2012machine,rossi2018mathematical}, seeking $\hat{\x}$ that maximizes the conditional probability $p(\y|\x)$, essentially maximizing the likelihood of the given measurements $\y$. Due to the Gaussianity of the noise, this probability is given easily by 
\begin{eqnarray}
p(\y|\x)=\const \cdot \exp\left\{ \frac{-\|\x-\y\|_2^2}{2\sigma^2}  \right\},
\end{eqnarray}
and maximizing it amounts to the trivial and fruitless solution: ${\hat \x}_{MLE}=\y$. This outcome is a direct manifestation of the ill-posedness of the denoising problem, exposing the need for more information for its solution. 

\begin{eqnarray}
p(\x|\y) = \frac{p(\y|\x)\cdot p(\x)}{p(\y)} = 
\const \cdot \exp\left\{ \frac{-\|\x-\y\|_2^2}{2\sigma^2}  \right\}\cdot p(\x).
\end{eqnarray}
In the last equality we have absorbed the denominator $p(\y)$ into the constant as it is independent of $\x$. While this expression is a simple modification to the MLE (multiplying the likelihood by the prior $p(\x)$), this is in fact a significant change, as it regularizes the inversion process from $\y$ to $\x$. 

Two commonly used estimators that exploit $p(\x|\y)$ are the MAP and the MMSE. The first is obtained by maximizing this posterior, leading to the Maximum A'Posteriori Probability (MAP) estimation~\cite{murphy2012machine,rossi2018mathematical}, given by\footnote{This minimization is obtained by taking the $-\log$ of the above expression.}  
\begin{eqnarray}
\label{eq:MAPest}
{\hat \x}_{MAP} = \argmin_{\x} \left[ \frac{\|\x-\y\|_2^2}{2\sigma^2}  - \log\left( p(\x) \right) \right].
\end{eqnarray}
As opposed to the MLE, ${\hat \x}_{MAP}$ is dictated by two forces, the first pulling it towards $\y$, while the other seeks a ``well-behaved'' result that leads to a low value of $-\log \left(p(\x )\right)$ -- this is exactly the regularization mentioned above. 

Similarly, the MMSE estimation~\cite{jaynes2003probability} is also reliant on the posterior probability obtained, as shown in Equation (\ref{eq:MMSE}), via\footnote{See Appendix \ref{App:MMSE} for a derivation of this statement.}
\begin{eqnarray}
\label{eq:MMSE2}
{\hat \x}_{MMSE} = \mathbb{E}\left( \x | \y \right) = \int_\x \x p(\x|\y) d\x.
\end{eqnarray}
While this expression is very concise and clear, operating with it has proven to be quite challenging due to the need for the partition function -- the normalizing factor of this distribution. 
This explains the vast popularity of the MAP-based approach among the classical methods. 

Be it the MAP or the MMSE, the Bayesian point of view requires access to $p(\x)$ or proxies of it. This brings us to the next discussion on the evolution of priors in image processing and their impact on the design of denoisers. 


\subsection{Evolution of Priors}
\label{sec:Evolution}

A key player in image processing is the prior, $p(\x)$, the probability density function of the image distribution. Modeling $p(\x)$ and using it for problems in visual data processing have served as the skeleton of our field, and defined its trajectory over the years. Below we outline the central milestones in the evolution of modeling $p(\x)$. 

One critical theme to remember is the fact that the expression $-\log (p(\x))$, which appears in the popular MAP estimation (see equation \eqref{eq:MAPest}), should assume a closed-form expression so as to lend itself to a manageable numerical optimization. For this reason, most attempts to characterize $p(\x) $ have chosen to use the Gibbs distribution form~\cite{jaynes2003probability}, $p(\x)=c\cdot \exp\{-\rho(\x)\}$, shifting our focus from $p(\x)$ to the energy function $\rho(\x)$. 

So, what should $\rho(\x)$ be to properly describe the image distribution? In order to keep this discussion concise, we present in Table \ref{tab:Priors} a brief list of possible analytical expressions for this function, without diving into their meaning, inter-relations, and effect. A more detailed explanation of these expressions is provided in Appendix \ref{App:Priors}. Please bear in mind the fact that this naïve approach of choosing an expression for $\rho(\x)$ is nothing short of a fantastic feat -- can we really expect a simple formula to grasp the richness of the image content distribution? 

The evolution of the ideas in Table \ref{tab:Priors} is characterized by several major and interconnected trends -- the migration from the familiar Gaussian distribution to the less intuitive heavy-tailed ones, the departure from $L_2$ to sparsity-promoting measures such as the $L_1$, the drift from linear approximation techniques (\textit{e.g.} PCA) to non-linear ones (\textit{e.g.} wavelets and sparse modeling), and above all, the replacement of axiomatic expressions with learned priors.

\setlength{\tabcolsep}{5pt}
\begin{table*}
\begin{center}
\caption{Evolution of priors for images.}
\label{tab:Priors}
\begin{tabular}{|l|l|l|}
\hline 
Years & Core concept & Formulae for $\rho(\cdot)$ \\
\hline
$\sim$ 1970 & Energy regularization & $\|\x\|_2^2$ \\
\hline
1975-1985 & Spatial smoothness & $ \|\L \x\|_2^2$ or $ \|\D_v \x\|_2^2 + \|\D_h \x\|_2^2$  \\
\hline
1980-1985 & Optimally Learned Transform & $\|\T \x\|_2^2 = \x^T \R^{-1} \x $  (via PCA)\\
\hline
1980-1990 & Weighted smoothness & $\|\L \x\|_{\W}^2$  \\
\hline
1990-2000 & Robust statistics & $\1^T \mu\{\L \x\} $ \textit{e.g.}, Hubber-Markov  \\
\hline
1992-2005 & Total-Variation & $\int_{v\in\Omega} |\nabla \x(v)| dv = \1^T \sqrt{|\D_v \x|^2 + |\D_h \x|^2}$  \\
\hline
1987-2005 & Other PDE-based options & $\int_{v\in\Omega} g\left[\nabla \x(v),\nabla^2 \x(v)\right] dv$ \\
\hline
2005-2009 & Field-of-Experts & $\sum_k \lambda_k \1^T  \mu_k \{\L_k \x\} $  \\
\hline
1993-2005 & Wavelet sparsity & $\|\W \x\|_1$  \\
\hline
2000-2010 & Self-similarity & 
$\sum_k \sum_{j\in \Omega(k)} d\{\R_k \x ,\R_j \x\}$ \\ 
\hline
2002-2012 & Sparsity methods & $\|\alpha\|_0 ~s.t.~ \x=\D\alpha$  \\
\hline
2010-2017 & Low-Rank assumption & $\sum_k \|\X_{\Omega(k)}\|_*$ \\
\hline
\end{tabular}
\end{center}
\end{table*}


\subsection{Other Classical Denoisers}
\label{sec:OtherDenoisers}

While the above-described Bayesian approach has proven to be quite productive, yielding a wide variety of denoising methods, alternative and more direct design techniques for such algorithms were also considered. Here we mention few such methods, some relying on the general notion of spatially adaptive smoothing of image content, while others leverage self-similarity that often-times exists in images. 

Consider the following rough motivating idea: Recall that a denoiser should attenuate random i.i.d. Gaussian noise while preserving the image content. When operating on a noisy pixel $y[i,j]$, our intuitive strategy is to open a neighborhood around it, $\Omega[i,j]$, for averaging purposes. If it so happens that the local image content in $\Omega[i,j]$ behaves like a tilted plane, a simple averaging of these neighborhood pixels would provide a perfect local noise suppression. When the local behavior deviates from this simple structure, the averaging mask should take this into account and adapt accordingly. 

This is exactly the idea behind the Bilateral filter~\cite{tomasi1998bilateral,elad2002origin} and the Beltrami-Flow~\cite{sochen1998general}, in which the averaging weight takes into account two forces -- (i) the proximity of the weighted pixel to the center of the neighborhood; and (ii) the proximity of this pixel's value to the center pixel's value, indicating its relevance to the averaging. Computing these weights for each pixel $y[i,j]$ and normalizing them to sum to one creates the local averaging kernel to apply. This way, if $\Omega[i,j]$ covers an edge between two regions, averaging will be restricted to the ``relevant'' pixels while discarding others. Non-Local-Means~\cite{buades2005non} takes this approach one step further by widening $\Omega[i,j]$ to a semi-local region, and by assessing pixels' relevance to the averaging by patch-matching instead of scalar value comparisons. This way we keep the spatially adaptive averaging concept, but robustify it and make it non-local. Kernel-regression~\cite{takeda2007kernel} is also a spatially adaptive averaging technique, but one that relies on a local parametric estimation of  the pixels' gray-values in $\Omega[i,j]$. A 2D Gaussian is fitted to the pixels in $\Omega[i,j]$, dictating its orientation and span, and this way offering a smoothing along edges instead of across ones. 

Another direct image denoising method that deserves our specific attention, especially due to its superior performance, is the BM3D algorithm~\cite{dabov2007image}. This technique relies on the expectation that 2D-DCT transformed local patches in natural images are expected to be sparse. Furthermore, by gathering groups of similar patches from the overall image area, this transformed sparsity should align in support. Thus, BM3D builds a 3D cube of similar patches for each pixel $y[i,j]$, transforms this cube together and forces a joint sparsity outcome. Among the classical denoising algorithms, BM3D is considered among the very best approaches in terms of MSE results. In this context, we also mention the Weighted Nuclear Norm Minimization (WNNM) denoising method~\cite{WNNM_2014} and its followups (\textit{e.g.}~\cite{yair2018multi}). These rely on a similar rationale to the BM3D, but replace the joint sparsity by a low-rank assumption.  


\subsection{Is denoising dead?}
\label{sec:Dead}

The paper ``Is Denoising Dead?'', published in 2009 by Chatterjee and Milanfar~\cite{PeymanDiD}, exposed a disturbing feeling shared by many in our community at the time -- a suspicion that we are touching the ceiling in terms of denoising ability. This impression relied on the considerable progress in design of denoising algorithms during the preceding years, and the fact that very different approaches towards this problem were found to lead to comparable denoising performance. A followup work~\cite{levin2011natural,levin2012patch} by Levin and Nadler in 2011-2012 addressed the same question. Both lines of work suggested a derivation of an approximated lower-bound of the MSE for noise removal ability. Without diving into the specifics of their derivations, we should mention that both concluded that there is still room for some improvement, even though this claim was not made constructively, leaving the question of how to obtain better techniques vague at best. 

From a practical point of view, and despite these optimistic conclusions, the progress in denoising performance after 2010-2011 was very slow and of diminishing returns. Indeed, the graph in Figure \ref{fig:Papers} shows a decrease in the number of papers on image denoising around 2010. However, this setback held true mostly for classically oriented methods of the kind discussed above. The emergence of deep neural networks brought a massive change to our domain, shattering the common belief about the end of this field, and the folklore around the attained performance limit. 

Indeed, deep learning brought new ways for the design of highly effective image denoisers, taking the lead in today's ability for noise suppression in images. However, the AI revolution had a much wider impact on the image denoising task, opening new horizons to possibilities and abilities never dealt with before. Among many such directions, these include (i) image adaptation; (ii) true noise removal; and (iii) addressing new denoising objectives. In the following section we discuss all these with much greater details. 

While the past decade can certainly be titled as the era of AI revolution, there has been another revolution, perhaps of a bigger scale, that took place in parallel in the field of image processing -- one that refers to the discovery that an image denoiser can serve other tasks. From the seminal paper on the Plug-and-Play Priors~\cite{PnP}, through Regularization by Denoising paper~\cite{RED}, and all the way to the recent and exciting diffusion-based image synthesis~\cite{song2019generative,ho2020denoising}, image denoisers are taking a new and much more exciting role in image processing. As this is the main theme of this paper, We shall expand on this line of work in Section \ref{sec:Discoveries} and after. 

So, to summarize, for the question `is denoising dead?' our answer is `definitely not!', and this is due to the vast influence of deep learning, and other new directions that brought new life to this domain. The rest of this paper is dedicated to the description of these developments and their impact and prospects. 


\section{Image Denoising -- The Deep Learning Revolution}
\label{sec:DL}

The recently discovered ability to effectively train deep neural networks for classification, regression and other tasks should not be taken lightly. Nothing in this process is well-understood or well-justified. Indeed, the opposite is true -- with overparametrized networks and a highly non-convex objective function, it is quite surprising that such networks are able to learn and generalize at all. And yet they do! This is the essence of the AI revolution that has found its way to so many fields, impacting each in a profound way. 

Image processing and computational imaging is yet another playground that has been deeply influenced by this AI revolution. Today's practice and theory in image processing is entirely different from the ones considered only 10 years ago. Indeed, image processing undergraduate and graduate courses had to change dramatically due to these new winds of change. 

And all this brings us to the new era of image denoising. In Section 
\ref{sec:Classic} we asked how should image denoisers be designed, and gave an answer that relies on the classical Bayesian approach. We now return to this question, and provide an entirely different answer -- one that builds on supervised deep-learning. This approach takes the following steps: 
\begin{enumerate}
    
    \item Start by gathering a large\footnote{By `large' we mean  thousands and sometimes millions of images, and the more the better. Often-times, the training itself may rely on several hundreds of images, and these are augmented by randomized operations such as crop, scale-down, rotations, and more.} dataset of clean images of diverse content - the kind of which we aim to denoise. We shall denote this set as ${\cal X}= \{ \x_k\}_{k=1}^M$. For simplicity assume that all images are of the same size. If this is not the case, an easy process of random tile extraction may convert the given data to this desired structure. 
    
    \item Recall that our goal is a design of a denoiser that removes additive white Gaussian noise of a specific strength $\sigma$. Thus, the next step is to create noisy instances of ${\cal X}$, \textit{i.e.} ${\cal Y}= \{\y_k\}_{k=1}^M$, where for $1\le k \le M, ~ \y_k= \x_k + \v_k$ and $\v_k \sim {\cal N}(0,\sigma^2 \I)$. In fact, every example $\x_k$ could be contaminated by several noise realizations, this way enriching the training set. 
    
    \item Define a parametric denoising architecture ${\hat \x}=D_\Theta(\y,\sigma)$ that should be trained to perform the denoising task. This stage is necessarily vague as there are many options for constructing such an architecture, and there seems to be no clear guidelines for its structure. Indeed, the literature offers various such options conceived by trial-and-error, \textit{e.g.}. More details and a discussion on this delicate stage is given below. 
    
    \item Define the training loss -- a penalty function that exploits the availability of ${\cal X}$ and ${\cal Y}$ and the defined parametric denoiser $D_\Theta(\y,\sigma)$, posing a cost value to be minimized with respect to $\Theta$, encouraging the denoised images to be close to their corresponding ideal ones. Such a functional could be of the form
    \begin{eqnarray}
    {\cal L}(\Theta) = \sum_{k=1}^M \dist\left(\x_k , {\hat \x}_k \right) = \sum_{k=1}^M \dist\left(\x_k , D_\Theta (\y_k,\sigma) \right),
    \end{eqnarray}
    where $\dist(\x,{\hat \x})$ is a distance function between the ideal and the denoised image, such as MSE -- $\dist(\x,{\hat \x})=\|\x-{\hat \x}\|_2^2$.
    
    \item Minimize ${\cal L}(\Theta)$ with respect to $\Theta$ via stochastic gradient descent~\cite{sgd} applied on small batches of training pairs $(\x_k,\y_k)$, and exploiting back-propagation~\cite{backprop}. 
    
\end{enumerate}

\noindent Once all the above steps are completed, the denoiser ${\hat \x}=D_\Theta(\y,\sigma)$ is ready to be deployed on newly incoming images, expected to perform better or worse in noise removal, depending on the size and quality of the training set, the similarity between the image to be denoised and the training set, the chosen architecture, and the quality and the hyperparameters of the optimization process. 

A variant of the above is blind denoising, in which $\sigma$ is unknown. The straightforward approach towards this task is brute-force learning. This means that for every ideal image $\x$ we produce a sequence of noisy versions $\y_k^\sigma$ with varying values of $\sigma$ in the range we aim to cover. Then learning is done by minimizing a loss that integrates over all the noise levels, 
\begin{eqnarray}
{\cal L}(\Theta) = \int_\sigma \sum_{k=1}^M dist\left(\x_k , D_\Theta (\y_k^\sigma) \right).
\end{eqnarray}
Observe that in this case the denoiser $D_\Theta$ gets only the noisy image without $\sigma$. An interesting alternative to the above was discovered in \cite{mohan2019robust}, showing that a bias-free architecture becomes robust to the noise power, and thus a simple training for a single value of $\sigma$ generalizes well to other levels of noise. 

An amazing consequence of all the above description is this: All the glorious work on image priors that fueled the design of classical denoisers and other tools in image processing seem to have become totally obsolete. Observe that in this supervised deep learning approach we have no need nor room for all the knowledge and know-how that have been accumulated carefully over decades of extensive research and engineering work. Is this a fair description of the current state of things in our field? To a large extent, the sad answer is positive, while some reservations to this conclusive statement will be discussed in Section \ref{sec:Synergy}. 

The emergence of deep learning techniques and their new abilities brought a new evolution of ideas on the design and span of image denoisers. While this literature is vast and rich, we describe below several key trends in this progress, in an attempt to expose both the new abilities obtained, and the new ideas accompanying them. These come in several fronts:

\begin{itemize}

\item {\bf Better Denoisers:} Improving image denoising capabilities via deep learning became a natural new front, where the aim is to perform better in terms of Peak-Signal-to-Noise-Ratio (PSNR) on an agreed-upon corpus of test images. This is manifested by an evolution of architectures that started with simple feed-forward Convolutional Neural Networks (CNN)~\cite{DnCNN}, proceeded to more advanced structures, such as UNet~\cite{unet, gurrola2021residual}, and all the way to the recently introduced Transformers~\cite{dosovitskiy2020image, liang2021swinir, zamir2022restormer}. In Figure \ref{fig:PSNrvsTime} we illustrate this trend by presenting a graph that shows the progress in denoising PSNR on the well-known BSD68 dataset~\cite{Martin2001ADO}. More details on each of these algorithms is brought in Appendix \ref{App:DenoisingAlgorithms}. 

\begin{figure}[htbp]
    \centering
    \includegraphics[width=\textwidth]{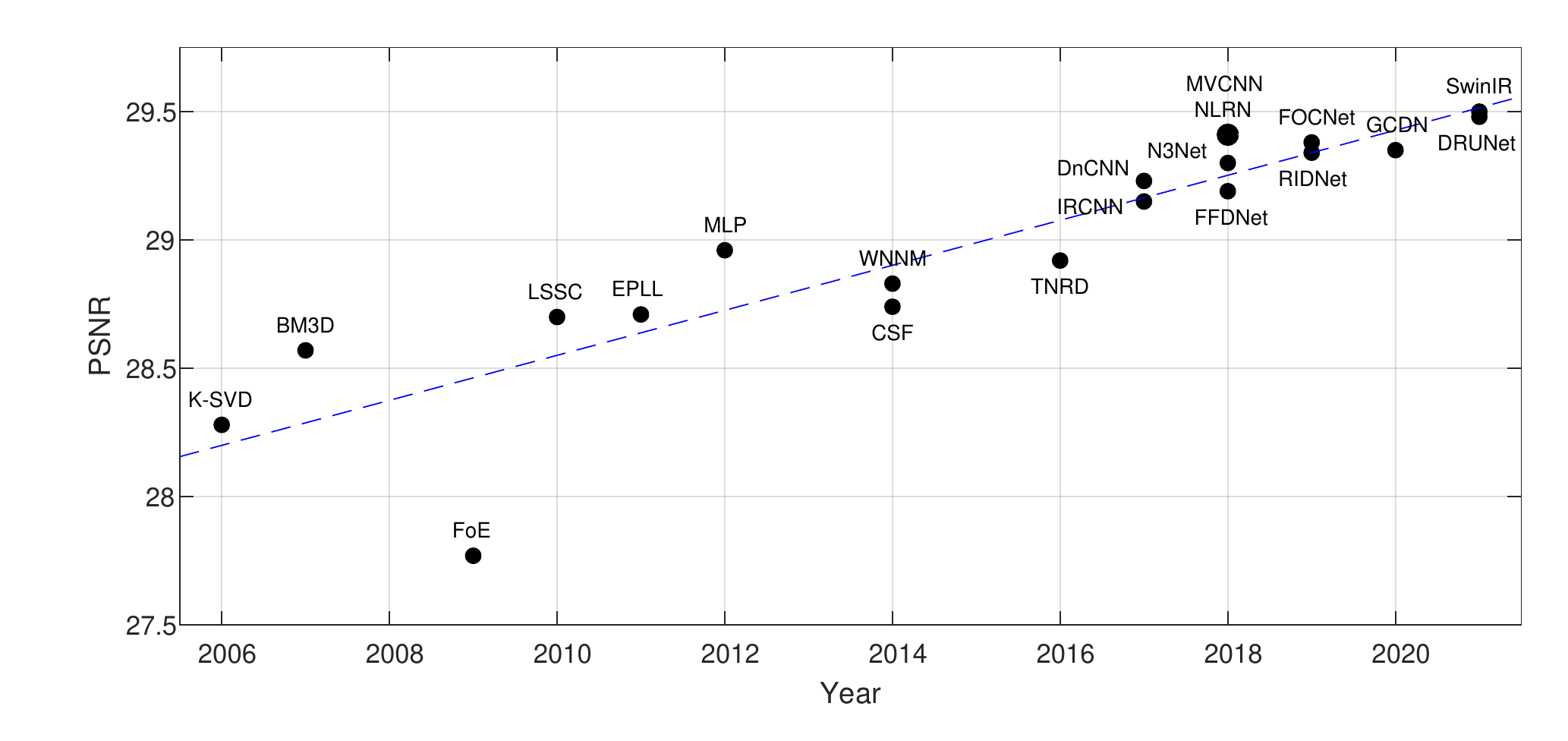}
    \caption{Denoising performance on the BSD68 dataset~\cite{Martin2001ADO} with $\sigma = 25$ (K-SVD~\cite{elad2006image}, BM3D~\cite{dabov2007image}, FoE~\cite{Roth2009FieldsOE}, LSSC~\cite{mairal2009non}, EPLL~\cite{zoran2011learning}, MLP~\cite{burger2012image}, CSF~\cite{Schmidt2014ShrinkageFF}, WNNM~\cite{WNNM_2014}, TNRD~\cite{chen2016trainable}, DnCNN~\cite{DnCNN}, IRCNN~\cite{zhang2017learning}, NLRN~\cite{liu2018non}, MVCNN~\cite{Liu2018MultilevelWF}, N3Net~\cite{Pltz2018NeuralNN}, FFDNet~\cite{zhang2018ffdnet}, FOCNet~\cite{Jia2019FOCNetAF}, RIDNet~\cite{anwar2019real}, GCDN~\cite{Valsesia2019DeepGI}, SwinIR~\cite{liang2021swinir}, DRUNet~\cite{Zhang2020PlugandPlayIR}).}
    \label{fig:PSNrvsTime}
\end{figure}

\item {\bf Different Training Schemes:} We described above the most obvious, supervised, training strategy, where we gather pairs of ideal images and their noisy version. Various unsupervised alternatives have been also developed for this task, such as Noise2Noise~\cite{noise2noise}, Noise2Void~\cite{noise2void}, Noise2Self~\cite{noise2self}, SURE-based denoising~\cite{zhang1998adaptive, luisier2007new, nguyen2020hyperspectral}, and others, all aim to operate on noisy images directly without the need for an access to their clean versions. It should be clear, though, that these techniques become relevant only in cases where the noise does not follow a known analytic structure, as otherwise the supervised alternative would be preferred. Another appealing approach that adopts an unsupervised denoiser training is ``Deep Image Prior'' (DIP)~\cite{ulyanov2018deep}, where a network is trained on a single image to best fit itself. An early stopping of this learning is shown to yield an effective denoising, revealing the regularization capabilities of the UNet architecture. 

\item {\bf True Noise Removal:} We mentioned above the Noise2X line of work~\cite{noise2noise, noise2void, noise2self}, which enables denoising of images without access to their clean versions. This ability becomes crucial when operating on images with un-modeled and unknown noise statistics. In such cases, learning should rely on more fundamental forces, such as self-similarity in images, the slow tendency of regressed neural networks to recreate noise from noise, the joint information that exists in burst of frames, and more. More broadly speaking, removal of true noise from images is a relatively new topic in image denoising, as it has hardly been addressed in the classical era due to its evident complexity. With advanced self-supervised and unsupervised learning techniques, new impressive abilities were created~\cite{wang2020practical, kim2014image, liu2021invertible, tran2020gan, wei2020physics, ignatov2021fast}. 

\item {\bf Image adaptation:} This refers to the ability to take an already designed/trained denoiser and adapt it to perform better on unique images that deviate from the training set. This way, general purpose denoisers could be boosted when operating on scanned documents, astronomical images, cartoon images and more. The adaptation itself could be done in various ways, the most natural of these is the following~\cite{vaksman2020lidia}: Given a noisy yet unique image to be cleaned, apply first the available denoiser $D_{\Theta_0}$ and obtain ${\hat \x}_0=D_{\Theta_0}(\y,\sigma)$. Now retrain the denoiser (\textit{i.e.} update the parameters $\Theta$) by minimizing $dist\left({\hat \x}_0, D_\Theta (\y,\sigma) \right)$. Similar to the core idea behind Noise2Noise~\cite{noise2noise} and DIP~\cite{ulyanov2018deep}, few gradient steps of this minimization are expected to go in the proper direction and yield a more informative and relevant denoiser, thus boosting the result for this specific image. The final outcome is obtained by ${\hat \x}=D_\Theta(\y,\sigma)$, using the slightly updated parameters $\Theta$.

\item {\bf Addressing Different Objectives:} When describing the supervised learning strategy of denoisers, we offered the $L_2$ loss that considers PSNR performance. Over the years this quality measure took the lead in most papers, despite its known weaknesses. Indeed, our community has been constantly striving to get the MMSE denoiser, if not in body, then at least in spirit, and this is evident from the PSNR performance tables that appear in almost every paper on image denoising published over the years. As we argue later on in Section \ref{sec:HPC-Denoising}, while MMSE denoisers are of great value by themselves, their outcome is not necessarily visually appealing, being an average over many potential solutions. 

Bearing this in mind, the learning paradigm creates a new opportunity for serving ``new masters'' -- recall that the learning loss function is highly non-convex, and yet we have no fear of its complexity when training the neural networks. Thus, we can easily replace the pleasant $L_2$ by more sophisticated or adequate penalties. The immediate alternative that comes to mind is SSIM~\cite{ssim}, which offers a more robust distance measure between images by considering structural similarity. We could go further and consider perceptual losses such as LPIPS~\cite{lpips}, that is further robustified by a learned representation in order to fairly assess proximity between images. This trend can be characterized as an attempt to produce visually pleasing and crisp images from the denoisers, ones that will surpass the MMSE alternative. A step forward in this direction takes us to Generative Adversarial Networks (GANs) for denoising~\cite{divakar2017image, dey2020image, ohayon2021high}. The idea is to challenge the output of the denoiser, by feeding it into a classifer that should tell apart true images versus denoised ones. By leveraging this classifier's guidance, the denoiser can learn to produce better looking images. We will come back to this idea in Section \ref{sec:HPC-Denoising}, offering an improved approach that targets perfect perceptual quality results. 

\end{itemize}

\noindent The description given above provides nothing but a glimpse into a very vibrant and rich body of literature that finds image denoising as an appealing playground for research. Still, we stop the survey of deep learning for denoising here, as our prime goal is the denoisers themselves and algorithms building on top of them. 

As one final note, observe that all the preceding discussion on classical and modern denoisers' design is given without referring to color images. Indeed, the formulation in this paper considers a grayscale image $\x$, yet most denoisers, old and new, are typically required to process color (Red-Green-Blue) images. Some of the existing methods discussed above are easily extended to color by operating on the three chroma channels jointly. For example, NLM~\cite{buades2005non} and K-SVD denoising~\cite{elad2006image} operate on RGB patches directly by flattening them to longer vectors. Another approach is to turn to the YUV or YCbCr color-space, and operate on the luma (Y) and the chroma (Cb/Cr or U/V) layers independently, as BM3D does~\cite{dabov2007image}. Denoisers based on deep neural networks typically handle color directly by feeding the RGB image as a 3-dimensional tensor input to the network, processed by subsequent 3D convolutions. More intricate approaches do exist, in which the geometrical interplay between the color layers is taken into account more adequately~\cite{sochen1998general}.


\section{Synergy between Classics and Deep Learning}
\label{sec:Synergy}

With the description given above, the reader may (rightfully!) get the impression that the vast knowledge regarding image denoising gathered during the classical era has become obsolete with the emergence of the deep learning alternatives. However, this claim is not entirely correct. In reality, the themes investigated and promoted by classical algorithms are serving as the foundations for building DL denoisers, even if practiced implicitly, and these are mostly manifested by the choice of architectures to be used. To illustrate this, we mention several well-known key concepts of classical image denoising algorithms, and show their impact on DL architectures:
\begin{itemize}
    \item \textbf{Locality:} Most information relevant to restoring a pixel's value in denoising is contained in its local neighborhood. In classical algorithms, this concept is embodied using patch processing, local filtering, local image priors, and more. When it comes to DL schemes, many denoisers choose convolutional layers as their primary processing path, which leads to architectures with small to moderate receptive fields~\cite{xu2015denoising, DnCNN, zhang2018ffdnet}.
    
    \item \textbf{Sparsity under appropriate transforms:} Local image patches are expected to be sparse when represented using certain 2D transforms. On the classical side, several of the priors listed in Table \ref{tab:Priors} fall into the sparsity-promoting category. On the DL side, a similar treatment can be observed, where the commonly used ReLU activation promotes sparsity by nulling the negatively activated neurons~\cite{relu}.
    
    \item \textbf{Self-similarity:} Most image patches have similar twins at other locations in the same image. While classical algorithms usually harness this property by gathering similar patches and processing them jointly, some recent DL schemes leverage self-similarity using self-attention layers~\cite{liang2021swinir, yao2022dense}.
    
\end{itemize}

\noindent Unfortunately, these and other concepts inherited from the classical era do not provide a constructive answer to the main question DL faces: How should we choose the appropriate architecture for the denoising task? Researchers facing this question are usually taking one of the two following options: (i) {\bf Copy:} adoption of an existing architecture that has been demonstrated to lead to good results in a similar task (\textit{e.g.}, DnCNN, UNet, ResNet, and more)~\cite{DnCNN,unet,resnet}. Usually such an adoption is accompanied by some minor modifications such as changing the number of channels or layers, etc.; or (ii) {\bf Trial and error:} gathering an architecture by piling a mix of known building blocks such as convolutions, strides, batch normalization steps, ReLU, fully-connected layers, down- and up-scaling, skip-connections, and more.

Both these options seem to work rather well, leading to networks achieving very good practical results -- see \cite{DnCNN, zhang2018ffdnet, liang2021swinir}. However, this brute-force approach typically tends to end up with heavy and cumbersome architectures, relying on millions of trainable parameters, making the resulting networks expensive to use and hard to train. Another downside in such architectures is their lack of explainability. While this may seem unimportant, having a black-box denoiser with no explainability implies an inability to leverage it to other tasks (\textit{e.g.}, image separation~\cite{miskin2000ensemble, fadili2009image, kutyniok2010image}), or probe it for identifying origins of failures for ill-treated regions in the image. More broadly, the brute-force approach towards architecture design for denoisers may require a lengthy trial and error process and may end up hitting a performance barrier.  

An alternative to copying or guessing architectures does appear in recent literature, known as \emph{unfolding}~\cite{gregor2010learning, zhang2020amp, scetbon2021deep, dutta2022deep, mou2022deep}. This approach constructs the neural network so as to mimic the computational stages of a well motivated algorithm. The term unfolding has to do with the fact that many classical image denoising methods involve iterative algorithms, and thus networks mimicking these should unfold their iterations to a feed-forward computational path. This approach typically produces concise and perfectly explainable networks, both in terms of the learned parameters and the activations obtained, which are easier to train. In addition, such networks tend to be easily and effectively adapted to different data. There are various examples in the literature for the unfolding approach for various regression tasks, \textit{e.g.}~\cite{yang2018proximal, corbineau2019learned, zhang2020deep, huang2020unfolding, vaksman2020lidia, scetbon2021deep, mou2022deep}. Here we briefly describe two such methods for illustrative purpose: Deep K-SVD~\cite{scetbon2021deep} and LIDIA~\cite{vaksman2020lidia}. Both propose a conversion of a classical denoising algorithm into a deep neural network architecture. 


\subsection{Deep K-SVD}
\label{sec:dksvd}

\begin{figure}[htbp]
    \centering
    \includegraphics[width=\textwidth]{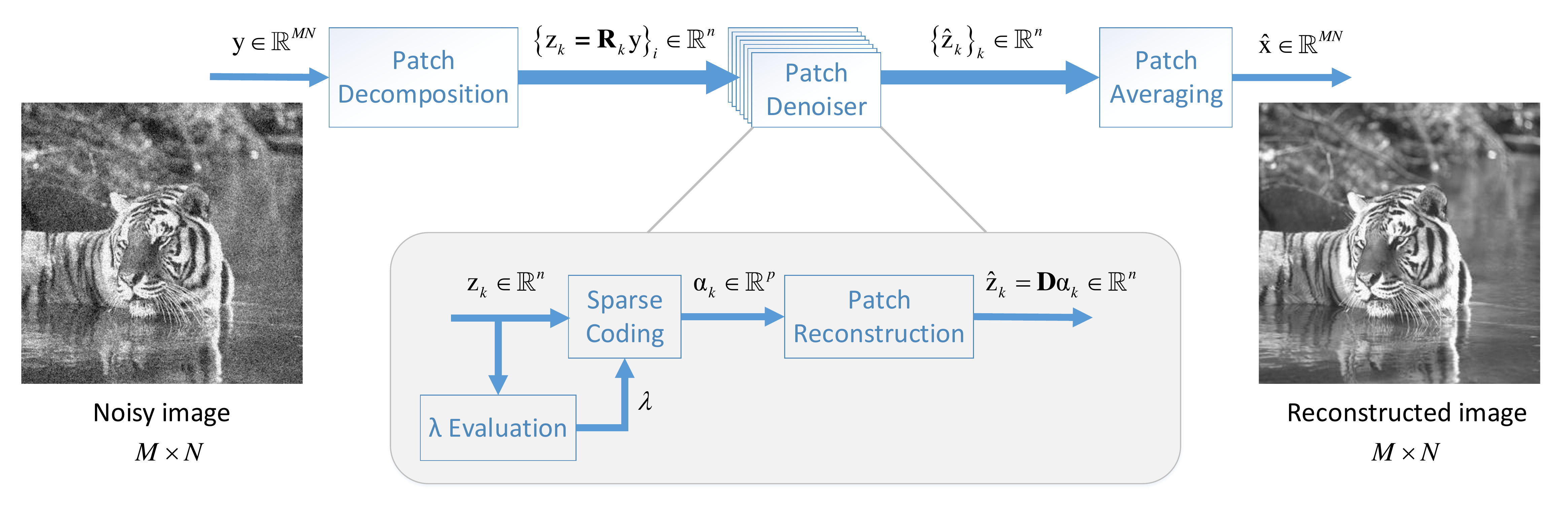}
    \caption{End-to-end architecture of the Deep K-SVD network~\cite{scetbon2021deep}.}
    \label{fig:dksvd}
\end{figure}
Deep K-SVD~\cite{scetbon2021deep} is an unfolding version of the K-SVD image denoising algorithm~\cite{elad2006image}. We start with a brief explanation of the original K-SVD method, and then turn to describe its unfolding. 

K-SVD denoising is based on sparse representation theory for constructing the image prior~\cite{aharon2006ksvd}. Consider a clean image $\mathbf{x}$ and patch extraction operators $\left\{\mathbf{R}_k\right\}_k$ such that $\mathbf{R}_k\mathbf{x} \in \mathbb{R}^{n}$ are image patches of size $\sqrt{n} \times \sqrt{n}$ taken from location $k$ in the image. The sparsity-promoting prior assumes that any such patch, $\mathbf{R}_k\mathbf{x}$, can be represented as a linear combination of {\emph few} columns (also referred to as atoms) from a redundant dictionary $\D \in \mathbb{R}^{n\times p}$ (redundancy implied by $p>n$), \textit{i.e.}, 
\begin{equation}
    \mathbf{R}_k\mathbf{x} = \D \alpha_k \;,
\end{equation}
where $\alpha_k \in \mathbb{R}^{p}$ is a sparse vector, $\left\|\alpha_i\right\|_0 \ll n$. Armed with this assumption, K-SVD poses the following minimization problem: 
\begin{equation}
    \min_{\left\{\alpha_k\right\}_k, \mathbf{x}} \quad \frac{\mu}{2} \left\|\mathbf{x} - \mathbf{y}\right\|_2^2 + \sum_k \left(\lambda_k  \left\|\alpha_k\right\|_0 + \frac{1}{2}\left\|\D\alpha_k - \mathbf{R}_k\mathbf{x}\right\|_2^2\right) \;,
\end{equation}
where $\mathbf{y}$ is the given noisy image, and $\mu$ and $\lambda_k$ are hyper-parameters. In this expression, the first term is the Log-Likelihood that requires a proximity between the reconstructed image $\mathbf{x}$ and the noisy image $\mathbf{y}$. The second and third terms represent the sparse representation prior, demanding that every image patch $\mathbf{R}_k\mathbf{x}$ in every location $k$ has an approximate sparse representation $\alpha_k$. 

The K-SVD algorithm solves this minimization problem by applying the following two steps iteratively: (i) Fix $\x$ (initialized by $\x = \y$) and update the vectors $\left\{\alpha_k\right\}_k$; and (ii) Update $\x$ while freezing the sparse representation vectors. The first is referred to as the sparse coding stage, where each patch in the contemporary solution obtains a sparse representation via the Orthogonal Matching Pursuit (OMP) greedy algorithm~\cite{pati1993orthogonal}. The second step becomes a quadratic minimization task, its solution being a simple variation of patch-based averaging. A single round of the above two steps has been shown to suffice for getting very good results~\cite{elad2006image}, and a repetition of this round several times could further boost the results~\cite{zoran2011learning}. The dictionary $\D$ in the above process could be either universal -- pretrained to best sparsify natural image content, or image adaptive -- updated to the image $\y$ itself within the above optimization. 

We now turn to describe the Deep K-SVD algorithm, which adopts the universal dictionary approach. The end-to-end architecture referring to a single round is illustrated in Figure~\ref{fig:dksvd}. This neural network consists of three main blocks: Patch Decomposition, Patch Denoiser, and Patch Averaging, all following closely the very same steps described above, with appropriate adaptations. Patch Decomposition breaks the input image $\mathbf{y}$ into a set of fully overlapped patches $\left\{\mathbf{z}_k\right\}_k = \left\{\mathbf{R}_k\mathbf{y}\right\}_k$. The next block, Patch Denoiser, is applied patch-wise, but replaces the OMP by LISTA~\cite{gregor2010learning}, in which  $\mathbf{z}_k$ undergoes sparse coding via a differentiable shrinkage-based iterative algorithm~\cite{gregor2010learning}. These inner iterations are unfolded as well to create a feed-forward computational path that starts with $\mathbf{z}_k$ and ends with $\mathbf{\hat z}_k = \D\alpha_k$. Due to the gap between OMP and LISTA, a sub-network of fully-connected layers computes the value of $\lambda_k$ for the incoming patch $\z_k$. The last block, Patch Averaging, rebuilds the reconstructed image $\mathbf{\hat{x}}$ by averaging the cleaned patches $\mathbf{\hat{z}}_k$ using learned weights. 

This Deep K-SVD network is trained end-to-end by minimizing the MSE distance between the ideal and denoised images for a set of $M$ training images,
\begin{equation}
\label{eq:mse_objective}
    \mathcal{L}\left(\Theta\right) = \sum_{k = 1}^M \left\|\mathbf{x}_k -\mathbf{\hat{x}}_k\right\|_2^2 = \sum_{k = 1}^M \left\|\mathbf{x}_k - D_{\Theta}\left(\mathbf{y}_k\right)\right\|_2^2 \;,
\end{equation}
where $\left\{\x_k,y_k \right\}_k$ is a set of clean and noisy image pairs to train on. $D_{\Theta}$ is the denoising network, where $\Theta$ stands for all trainable parameters, consisting of the dictionary $\D$, the parameters of the sub-network that evaluates $\lambda_k$ and the shrinkage thresholds.

Despite the close resemblance between the original algorithm and its unfolded version, the later performs much better~\cite{scetbon2021deep}, surpassing classical methods and aligning with deep-learning based techniques. This should not come as a surprise as the unfolded denoiser $D_{\Theta}$ is trained in a supervised manner, being fully aware of the task it serves, whereas the original algorithm relied on a ``guessed'' image prior. Interestingly, the universal dictionary obtained for $D_{\Theta}$ is markedly different from the one trained off-line for the original K-SVD denoising method, again a testimony to the major difference between the two design strategies.  


\subsection{LIDIA - Lightweight Learned Image Denoising}
\label{sec:lidia}

Another example of unfolding-based denoising is LIDIA~\cite{vaksman2020lidia}, which mimics the computational stages of the BM3D~\cite{dabov2007image}. As already mentioned in Section~\ref{sec:OtherDenoisers}, BM3D harnesses two prime forces for its denoising goal -- sparsity and self similarity. The first relies on the assumption that local image patches are sparse under the 2D-DCT spatial transform; the later is reflected by operating on groups of similar patches jointly, forcing sparsity again by transforming across these patches.

\begin{figure}[htbp]
    \centering
    \includegraphics[width=\textwidth]{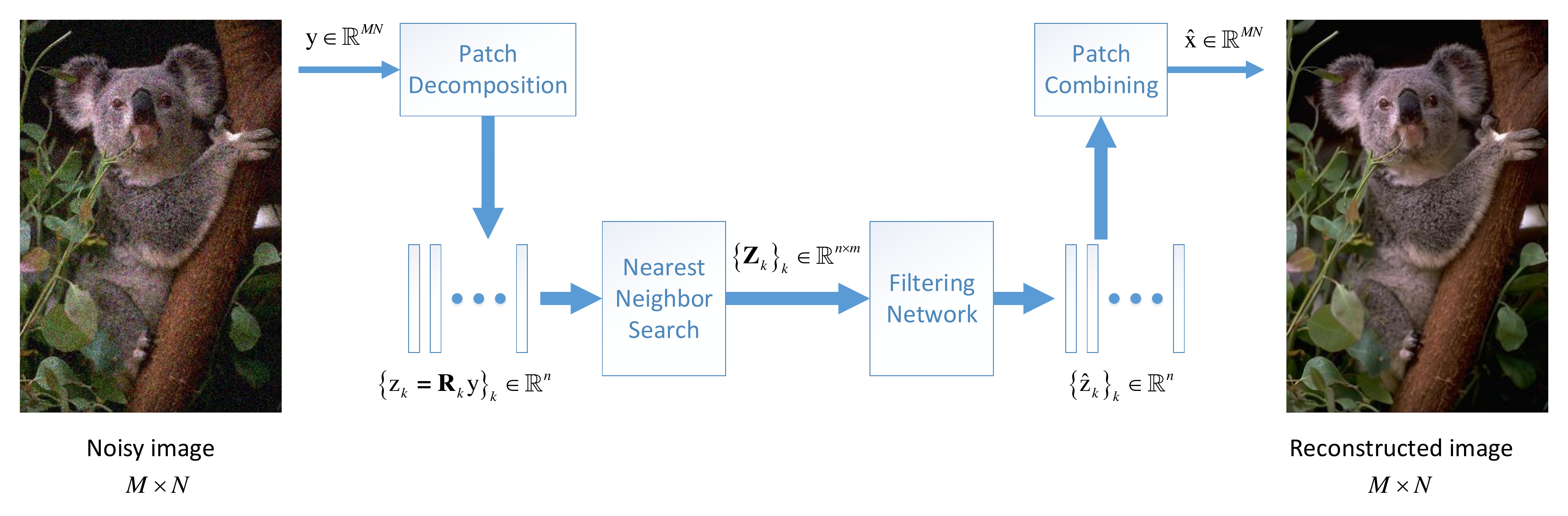}
    \caption{The LIDIA denoising computational path.}
    \label{fig:lidia}
\end{figure}
\begin{figure}
    \centering
    \includegraphics[width=\textwidth]{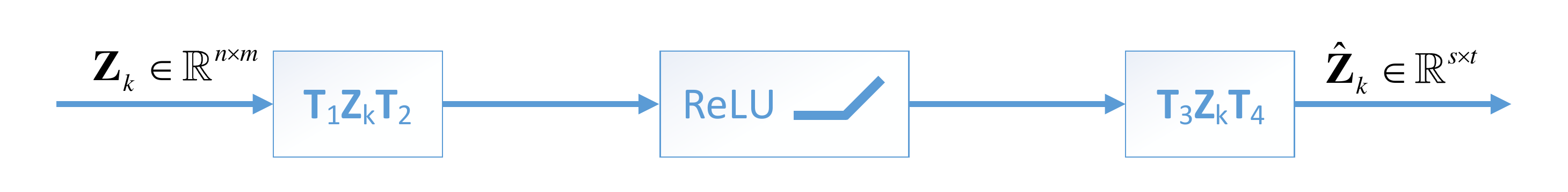}
    \caption{The Transform-ReLU-Transform block. Applying the matrices $\mathbf{T}_1$ and $\mathbf{T}_2$ transforms the input, $\mathbf{Z}_k$, to a space in which patches are supposed to be sparse; the matrices $\mathbf{T}_3$ and $\mathbf{T}_4$ transform the outcome to the pixel domain. Observe that the transform applied on $\mathbf{Z}_k$ is separable -- $\mathbf{T}_1$ is applied within patches while $\mathbf{T}_2$ operates across. This enables a reduction of the size of the matrices $\mathbf{T}_1, \dots, \mathbf{T}_4$ in order to enable their training.}
    \label{fig:trt}
\end{figure}

LIDIA's core computational path is shown schematically in Figure~\ref{fig:lidia}. This neural network starts by breaking the input image $\mathbf{y}$ into a set of fully overlapping patches $\left\{\mathbf{z}_k\right\}_k$ of size $\sqrt{n} \times \sqrt{n}$. Then, each patch, $\mathbf{z}_k \in \mathbb{R}^n$, is augmented with a group of its $m - 1$ nearest neighbors, forming a matrix $\mathbf{Z}_k$ of size $n \times m$. The filtering is applied patch-wise -- each matrix, $\mathbf{Z}_k$, undergoes a series of blocks composed of a separable transform, ReLU, and another separable transform, as shown schematically in Figure~\ref{fig:trt}. This mimics the BM3D operation by transforming the input matrix to a space in which local patches are believed to be sparse, forcing sparsity using the ReLU layer, and transforming back the outcome to the pixel domain. Unlike BM3D, the transforms are trainable and are not restricted to be the inverse of each other, nor forced to be square matrices. In addition, LIDIA includes a multi-scale treatment, simultaneously processing patches in several scales. During processing, the corresponding patches from different scales are fused using a learned joint transform. Finally, the reconstructed image is obtained by returning the denoised patches to their original places while averaging overlaps using learned weights. 

The LIDIA network is trained end-to-end (excluding the nearest-neighbor part) by minimizing the MSE loss, similar to the loss in Equation~\ref{eq:mse_objective}, applied on a set of $M$ training images. The network can be trained for a specific noise level $\sigma$ or blindly, aiming to serve a range of $\sigma$ values. LIDIA performs much better than the original BM3D algorithm since it uses learned rather than fixed transforms. Compared with other deep-learning techniques LIDIA achieves comparable results, while using a small fraction of the typical number of learned parameters. 

\begin{figure}[htbp]
    \centering
    \begin{subfigure}{0.24\textwidth}
        \includegraphics[width=\textwidth]{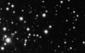}
        \caption{Clean }
    \end{subfigure}
    \begin{subfigure}{0.24\textwidth}
        \includegraphics[width=\textwidth]{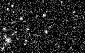}
        \caption{Noisy, ${\sigma = 50}$ }
    \end{subfigure}
    \begin{subfigure}{0.24\textwidth}
        \includegraphics[width=\textwidth]{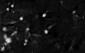}
        \caption{Denoised, $24.22$dB }
    \end{subfigure}
    \begin{subfigure}{0.24\textwidth}
        \includegraphics[width=\textwidth]{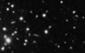}
        \caption{Adapted, $26.34$dB}
    \end{subfigure} \\
    \begin{subfigure}{0.24\textwidth}
        \frame{\includegraphics[width=\textwidth]{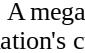}}
        \caption{Clean }
    \end{subfigure}
    \begin{subfigure}{0.24\textwidth}
        \frame{\includegraphics[width=\textwidth]{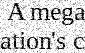}}
        \caption{Noisy, ${\sigma = 50}$ }
    \end{subfigure}
    \begin{subfigure}{0.24\textwidth}
        \frame{\includegraphics[width=\textwidth]{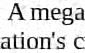}}
        \caption{Denoised, $22.10$dB }
    \end{subfigure}
    \begin{subfigure}{0.24\textwidth}
        \frame{\includegraphics[width=\textwidth]{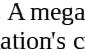}}
        \caption{Adapted, $25.82$dB}
    \end{subfigure}
    \caption{Image adaptation via LIDIA: The original denoising network is trained for general content images, and performs reasonably well for astronomy and scanned document inputs. A substantial boost in denoising performance is obtained for these two examples, due to their deviation from the training set.}
    \label{fig:ImageAdaptation}
\end{figure}

In Section \ref{sec:DL} we mentioned the ability to adapt a given denoiser to newly coming images that deviate from the training set. This adaptation starts by applying the trained denoiser, and then uses the output in order to fine-tune the denoiser parameters by applying few gradient steps. This rationale has been successfully demonstrated with LIDIA, and two such illustrative results are brought in Figure \ref{fig:ImageAdaptation}.  

\subsection{Summary - The classics is still here}

We described two unfolding instances in which classic denoising algorithms provide their architecture for the learned network. These and other such methods~\cite{zhang2020amp, scetbon2021deep, dutta2022deep, mou2022deep}, targeting various image recovery tasks, offer a constructive path towards well-motivated, low complexity and explainable neural architectures. In the quest for a synergy between classical denoising methods and novel deep-learning alternatives, this is probably the most natural manifestation of it. 


\section{Image Denoising -- Migration towards Recent Discoveries}
\label{sec:Discoveries}

\begin{figure}[htbp!]
    \centering
    \begin{subfigure}{0.32\textwidth}
        \frame{\includegraphics[width=\textwidth,trim={0 180px 0 76px},clip]{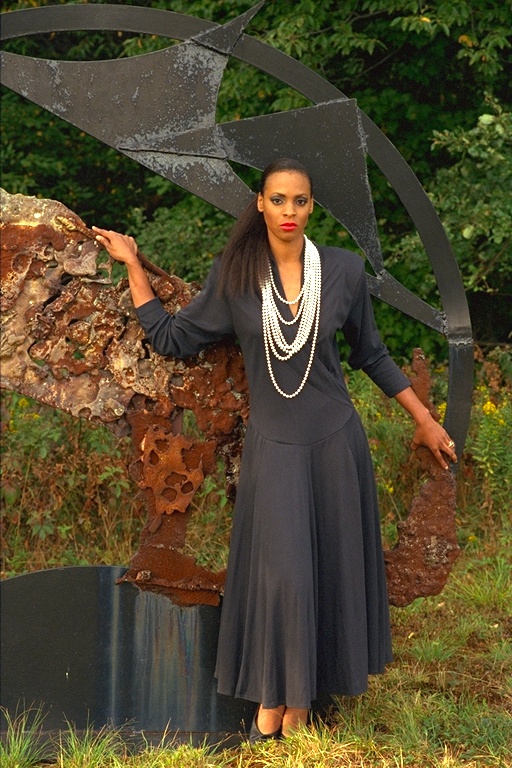}}
        \caption{Clean }
    \end{subfigure}
    \begin{subfigure}{0.32\textwidth}
        \frame{\includegraphics[width=\textwidth,trim={0 180px 0 76px},clip]{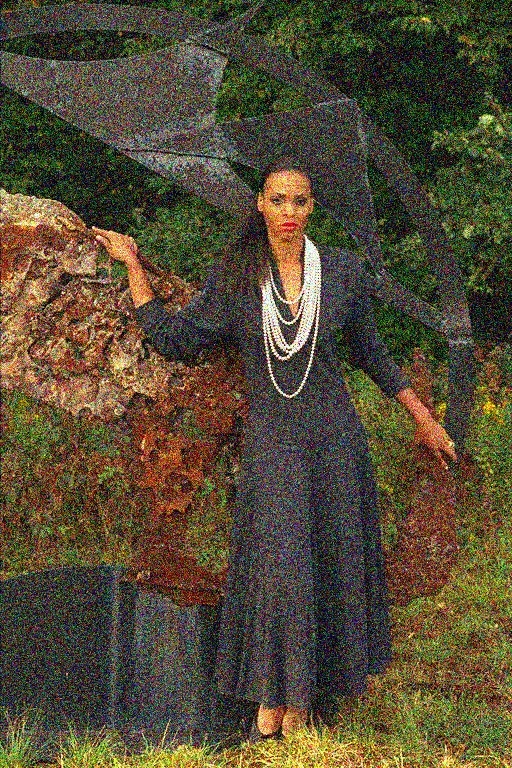}}
        \caption{Noisy, ${\sigma = 50}$ }
    \end{subfigure}
    \begin{subfigure}{0.32\textwidth}
        \frame{\includegraphics[width=\textwidth,trim={0 180px 0 76px},clip]{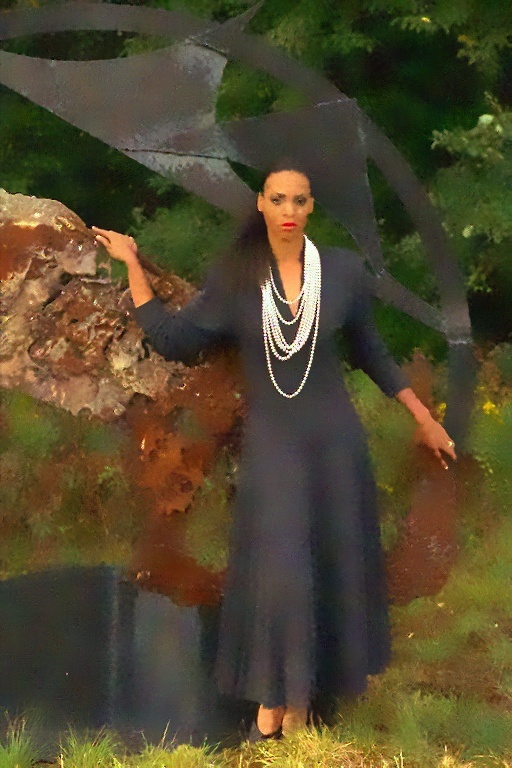}}
        \caption{NLM, $24.67$dB }
    \end{subfigure}
    \begin{subfigure}{0.32\textwidth}
        \frame{\includegraphics[width=\textwidth,trim={0 180px 0 76px},clip]{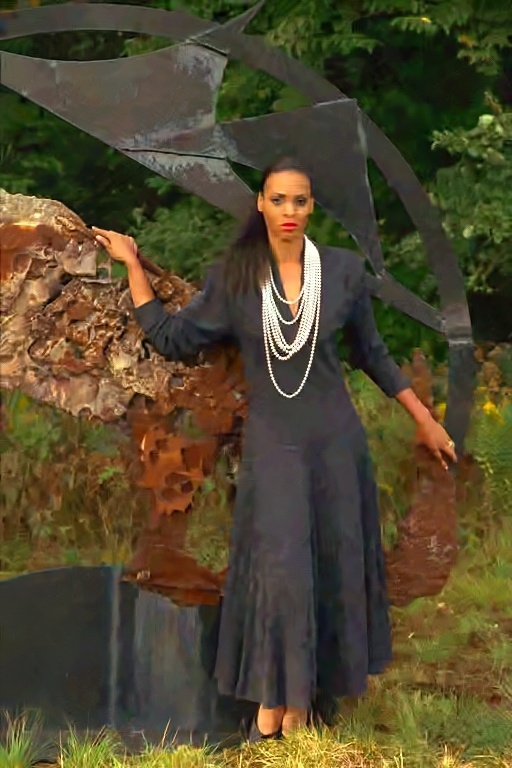}}
        \caption{BM3D, $26.31$dB }
    \end{subfigure}
    \begin{subfigure}{0.32\textwidth}
        \frame{\includegraphics[width=\textwidth,trim={0 180px 0 76px},clip]{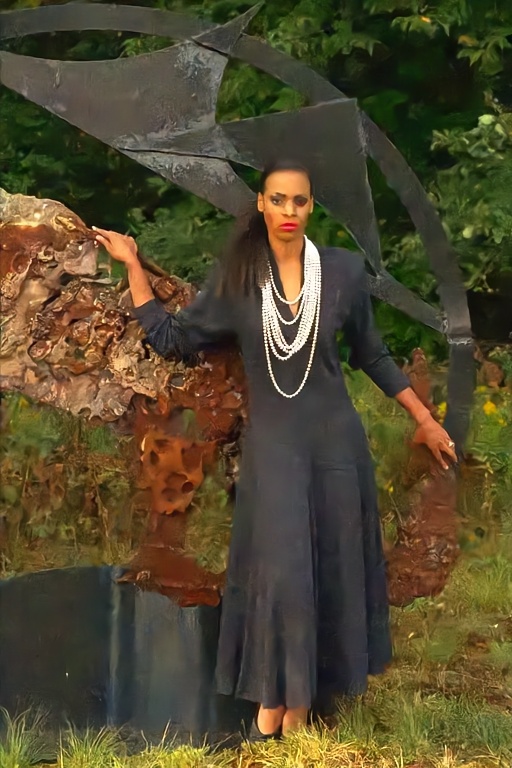}}
        \caption{DnCNN, $26.70$dB }
    \end{subfigure}
    \begin{subfigure}{0.32\textwidth}
        \frame{\includegraphics[width=\textwidth,trim={0 180px 0 76px},clip]{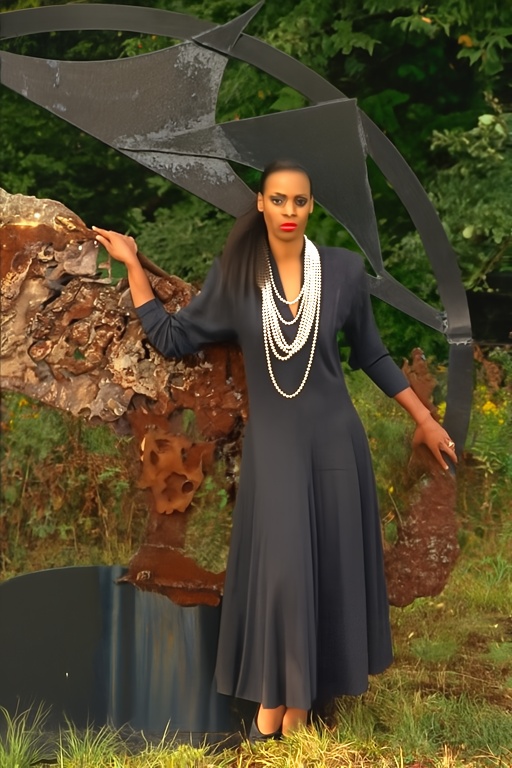}}
        \caption{SwinIR, $27.31$dB }
    \end{subfigure}
    \caption{Demonstration (1) of several denoising methods:  (NLM~\cite{buades2005non}, BM3D~\cite{dabov2007image}, DnCNN~\cite{DnCNN}, SwinIR~\cite{liang2021swinir}).}
    \label{fig:DenoisingResults1}
\end{figure}

\begin{figure}[htbp!]
    \centering
    \begin{subfigure}{0.32\textwidth}
        \frame{\includegraphics[width=\textwidth,trim={0 76px 0 180px},clip]{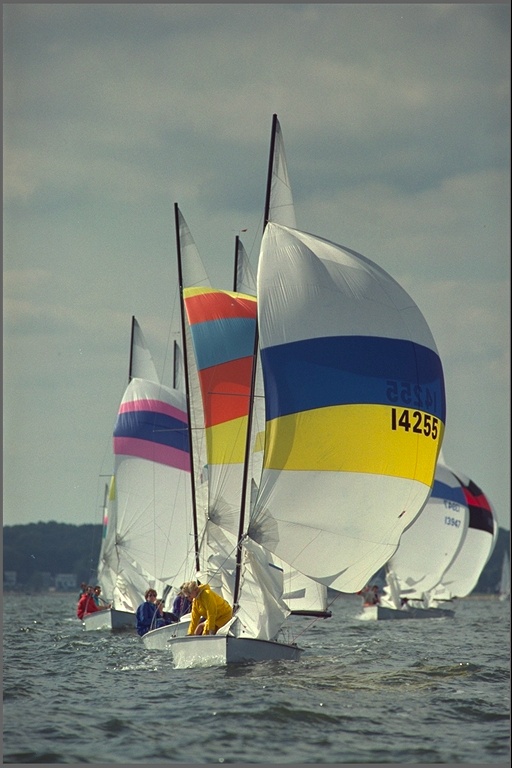}}
        \caption{Clean }
    \end{subfigure}
    \begin{subfigure}{0.32\textwidth}
        \frame{\includegraphics[width=\textwidth,trim={0 76px 0 180px},clip]{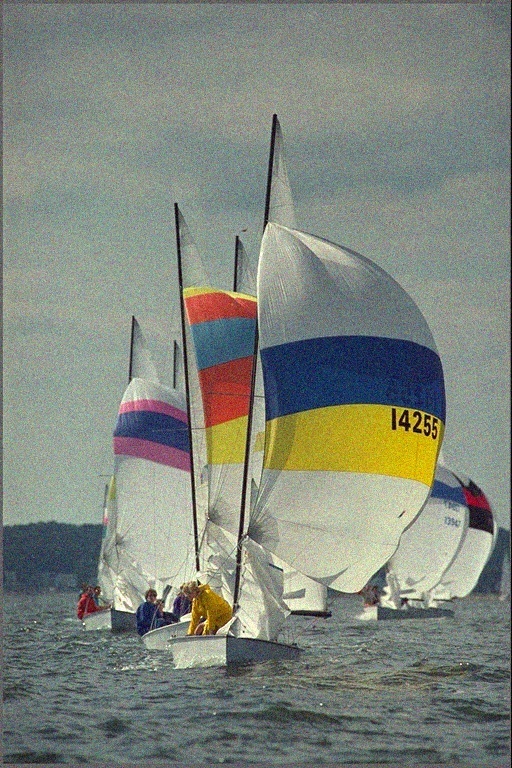}}
        \caption{Noisy, ${\sigma = 15}$ }
    \end{subfigure}
    \begin{subfigure}{0.32\textwidth}
        \frame{\includegraphics[width=\textwidth,trim={0 76px 0 180px},clip]{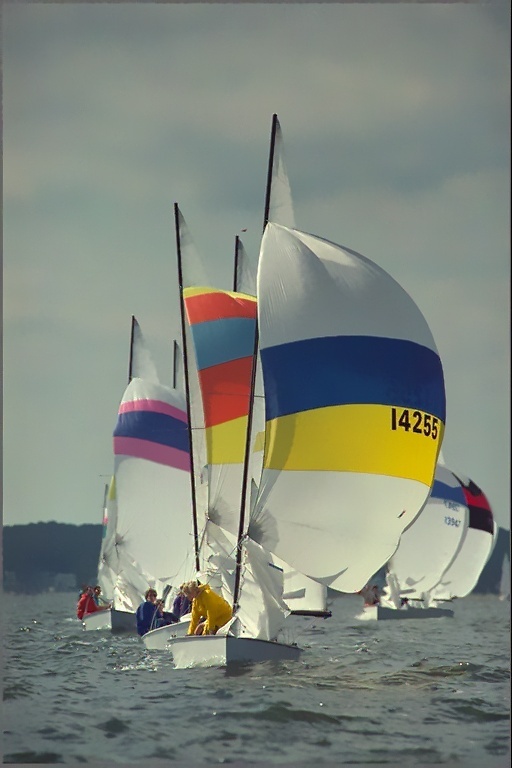}}
        \caption{NLM, $33.82$dB }
    \end{subfigure}
    \begin{subfigure}{0.32\textwidth}
        \frame{\includegraphics[width=\textwidth,trim={0 76px 0 180px},clip]{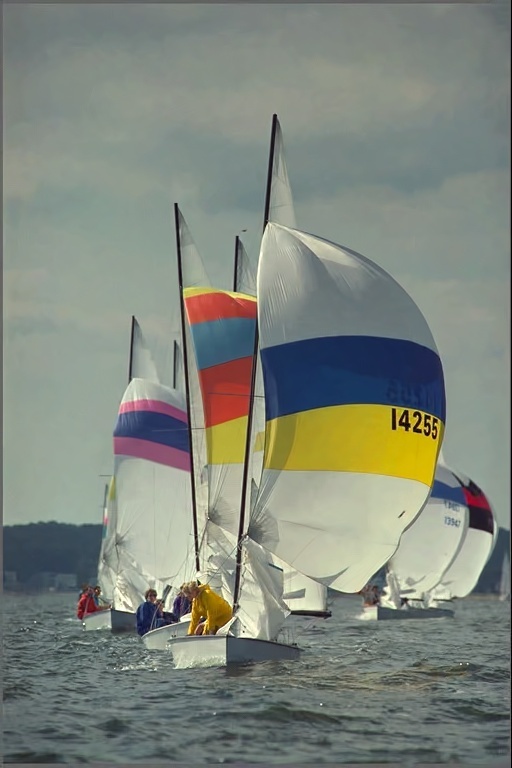}}
        \caption{BM3D, $36.23$dB }
    \end{subfigure}
    \begin{subfigure}{0.32\textwidth}
        \frame{\includegraphics[width=\textwidth,trim={0 76px 0 180px},clip]{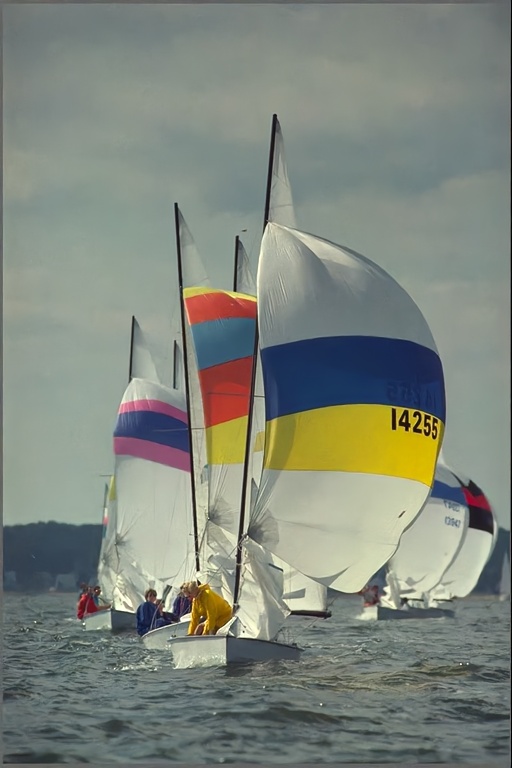}}
        \caption{DnCNN, $36.33$dB }
    \end{subfigure}
    \begin{subfigure}{0.32\textwidth}
        \frame{\includegraphics[width=\textwidth,trim={0 76px 0 180px},clip]{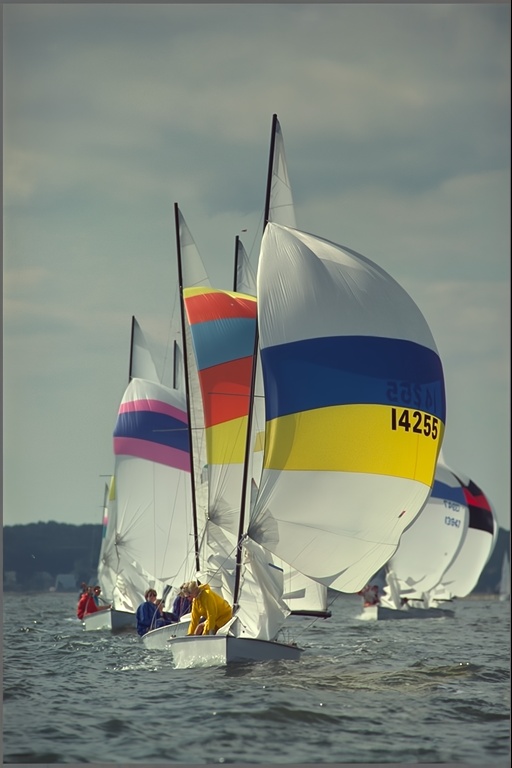}}
        \caption{SwinIR, $37.17$dB }
    \end{subfigure}
    \caption{Demonstration (2) of several denoising methods:  (NLM~\cite{buades2005non}, BM3D~\cite{dabov2007image}, DnCNN~\cite{DnCNN}, SwinIR~\cite{liang2021swinir}).}
    \label{fig:DenoisingResults2}
\end{figure}

The clear conclusions from the above discussion are these: Highly effective image denoisers for AWGN removal, $D(\y,\sigma)$, are definitely within reach, and the better ones are likely to be deep-learning based algorithms. In an attempt to illustrate these statements, Figures \ref{fig:DenoisingResults1} and \ref{fig:DenoisingResults2} present denoising results for two test images, two noise levels ($\sigma=15,50$) and by several denoisers -- NLM~\cite{buades2005non}, BM3D~\cite{dabov2007image}, DnCNN~\cite{DnCNN}, and SwinIR (a transformer-based denoising network)~\cite{liang2021swinir}. As can be seen, the results are very impressive and more so by the later deep neural network solutions. 

We now turn to ask far more daring questions with regard to such denoisers, focusing this time on their deployment to other tasks. More specifically, we discuss three such questions, each corresponding to a recent discovery in the field of imaging sciences:  
\begin{itemize}
    \item {\bf Discovery 1:} Can we leverage a denoiser $D(\y,\sigma)$ for solving general linear inverse problems? As we shall shortly see, the answer to this question is positive and constructive, opening new horizons for design of recovery algorithms and their  regularization. 
    
    \item {\bf Discovery 2:} Can we leverage a denoiser $D(\y,\sigma)$ for synthesizing (hallucinating) high-quality images, fairly drawn from the prior probability density function $p(\x)$? Here again the answer is positive and constructive, offering a thrilling new line of activity in machine learning. 
    
    \item {\bf Discovery 3:} If hallucination of perfect-looking images is achievable, can we revisit the topic of general linear inverse problems and leverage a denoiser $D(\y,\sigma)$ for their solution while targeting \emph{perfect perceptual quality} results? Here again we give a positive answer, and lead to a new and inspiring branch of research in inverse problems, offering novel view of their treatment. 
         
\end{itemize}

\noindent Below we discuss each of these discoveries in greater details. It is our sincere belief that these together form one of the most exciting eras for our field, marking a major transition in how image processing is perceived and practiced. 


\section{Discovery 1: Solving Inverse Problems via Image Denoisers}
\label{sec:IP}

Given a denoiser $D(\y,\sigma): \mathbb{R}^N\rightarrow \mathbb{R}^N$, our goal is to use it somehow for solving general linear inverse problems of the  form 
\begin{eqnarray}
\y = \H \x+\v,
\end{eqnarray}
where $\H \in \mathbb{R}^{M\times N}$ is a known matrix, $\v \in \mathbb{R}^M$ is AWGN, and $\y\in \mathbb{R}^M$ is the given measurement vector. Observe that $\H=\I$ stands for the denoising problem. Therefore, the current discussion extends our view to a wider family of tasks in imaging sciences, covering applications such as deblurring, inpainting, demosaicing, super-resolution, tomographic reconstruction, compressed sensing, and more. 

Following the derivations in Section \ref{sec:Classic} and specifically Equation (\ref{eq:MAPest}), we can adopt the Bayesian point of view and obtain the MAP estimation for this family of problems:   
\begin{eqnarray}
\label{eq:MAPestH1}
{\hat \x}_{MAP} = \argmin_{\x} \left[ \frac{\|\H\x-\y\|_2^2}{2\sigma^2}  - \log\left( p(\x) \right) \right].
\end{eqnarray}
Plugging in the Gibbs distribution form for the prior, $p(\x)\sim \exp\{- \rho(\x)\}$, this becomes
\begin{eqnarray}
\label{eq:MAPestH2}
{\hat \x}_{MAP} = \argmin_{\x} \left[ \|\H\x-\y\|_2^2  + c \cdot \rho(\x) \right].
\end{eqnarray}
Clearly, the greatest riddle posed above has to do with the identity of the energy function $\rho(\x)$. Can a denoiser serve all linear inverse problems in a unified approach by providing a connection or an alternative to $\rho(\x)$? Surprisingly, the answer to this question is positive and constructive. The seminal Plug-and-Play Prior (PnP) work by Venkatakrishnan, Bouman and Wohlberg~\cite{PnP} was the first to provide such an answer\footnote{We should note that an alternative, yet closely related, derivation is offered in~\cite{metzler2016denoising} from an approximate message passing point of view.}, followed and improved upon by RED (Regularization by Denoising)~\cite{RED}. These and their various extensions and variations have created a vivid and stimulating sub-field of research in imaging sciences~\cite{brifman2016turning,kamilov2017plug,tirer2018image,buzzard2018plug,sun2019online,chan2019performance,mataev2019deepred,teodoro2019image,ahmad2019plug,chen2021deep,cohen2021has} in which denoisers play a central role. Below we describe PnP and RED in more detail, and then turn to describe another, perhaps better founded, bridge between denoisers and the energy function $\rho(\x)$ via the \emph{score function}. This would serve our next step towards diffusion models, as they unravel in Section \ref{sec:Synthesis} and beyond. 


\subsection{Plug-and-Play Prior (PnP)} 

PnP~\cite{PnP} suggests the following steps in handling the minimization of the problem posed in Equation (\ref{eq:MAPestH2}): We start by splitting the variable $\x$ by defining $\z=\x$ and expressing each of the two penalties with a different variable: 
\begin{eqnarray}
\label{eq:PnP-Step1}
{\hat \x}_{MAP} = \argmin_{\x,\z} \left[ \|\H\x-\y\|_2^2  + c \cdot \rho(\z) \right] ~~ \subjectto ~~\z=\x.
\end{eqnarray}
The next step forms the Augmented Lagrangian of the above problem, converting the constraint into a penalty, 
\begin{eqnarray}
\label{eq:PnP-Step2}
L(\x,\z,\u) =  \|\H\x-\y\|_2^2  + c \cdot \rho(\z) +\lambda \|\z - \x + \u\|_2^2 -\lambda \|\u\|_2^2,
\end{eqnarray}
where $\u$ is the scaled dual variable and $\lambda$ is an (arbitrary) penalty weight (see more in \cite{PnP}). The third and final step applies ADMM~\cite{ADMM} for the minimization of $L(\x,\z,\u)$ with respect to $\x$ and $\z$ while updating $\u$. These  are obtained by alternating between the treatment of each variable while fixing the others: 
\begin{eqnarray}
\label{eq:PnP-Step3}
\x & \leftarrow & \argmin_{\x} \left[\|\H\x-\y\|_2^2 +\lambda \|\z - \x + \u\|_2^2 \right] = \left[\H^T\H+\lambda \I\right]^{-1}\left[\H^T \y+ \z-\u\right], \\
\z & \leftarrow & \argmin_{\z}  \left[ c \cdot \rho(\z) +\lambda \|\z - \x + \u\|_2^2  \right], \\
\u & \leftarrow & \u+(\x-\z).
\end{eqnarray}
In the above, the first update equation amounts to a simple Least-Squares, which does not involve $\rho(\x)$. The true drama takes place in the second update formula -- observe its close resemblance to Equation (\ref{eq:MAPest}), which formulates an image denoising task. Indeed, instead of choosing/guessing/learning $\rho(\x)$, we can apply our favorite denoiser ${\hat \z }=D(\x-\u,\sigma_0)$ where $\sigma_0$ should be inversely proportional to $\lambda/c$. This way, PnP offers an appealing iterative algorithm that repeatedly applies a denoiser in order to handle any underlying inverse problem, just as promised. 

While the original PnP paper did not dive into the issue of convergence of the above ADMM algorithm, nor posed conditions on the denoiser to support such guarantees, later work offers such a theoretical discussion -- we refer the interested readers to \cite{chan2016plug, xu2020provable, sun2021scalable, laumont2021bayesian}.


\subsection{Regularization by Denoising (RED)}

An alternative angle towards the relationship between $\rho(\x)$ and image denoising is presented in \cite{RED}. The core idea is quite simple, using the following explicit formula for $\rho(\x)$ that relies on a denoiser:
\begin{eqnarray}
\label{eq:rho-RED}
\rho(\x) = \x^T \left[ \x - D(\x,\sigma_0) \right].
\end{eqnarray}
The intuition behind this expression can be uncovered by considering a linearized form of the denoising process, $D(\x,\sigma_0) = S(\x)\x$, where $S(\x)$ is an image-dependent matrix that represents the smoothing applied by the noise removal process. This way, the chosen energy function becomes $\rho(\x) = \x^T \left[ \I - S(\x) \right]\x$, which is a Laplacian smoothness prior of the kind described in Section \ref{sec:Classic}, although being image-adaptive (and thus far more effective).  

The work in \cite{RED} shows that if the denoiser $D(\x,\sigma_0)$ is differentiable, passive and of symmetric Jacobian, the chosen energy function in Equation (\ref{eq:rho-RED}) is guaranteed to be convex. If, in addition, the denoiser satisfies a local homogeneity property\footnote{See \cite{RED} for the exact definitions of these ingredients and for the proof of their implications.}, then the following relationship holds: 
\begin{eqnarray}
\label{eq:Gradrho-RED}
\nabla_{\x} \rho(\x) = 2\left[\x - D(\x,\sigma_0)\right].
\end{eqnarray}
This relationship is a centerpiece in the construction of several RED algorithms. Plugging the chosen $\rho(\x)$ from Equation (\ref{eq:rho-RED}) into Equation (\ref{eq:MAPestH2}) implies that the gradient of this functional is easily accessible, requiring a single activation of the chosen denoiser. Critically, this gradient does not require the differentiation of $D(\x,\sigma_0)$, which would have required far more computational power and memory consumption. As a consequence, various gradient-based optimization strategies can be applied for computing ${\hat \x}_{MAP}$, and all are guaranteed to converge to the global minimizer of the MAP penalty. Again, we arrive at iterative algorithms that apply simple linear operations and a denoiser in each step, aiming to solve general linear inverse problems. 

An intriguing question with respect to the above is the identity of the denoiser to use within RED. Should it be an MMSE denoiser? Should it be designed to remove AWGN? Would these choices lead to the required properties mentioned above (diffentiability, symmetry, passivity, homogeneity)? What should $\sigma_0$ be? Partial answers to these questions are given by the next discussion on the \emph{score function}. 


\subsection{The Score Function and its Relevance to Inverse Problems}

Embarking from Equations (\ref{eq:MAPestH1}) and (\ref{eq:MAPestH2}), we now present a very different approach towards getting to the same RED formulation, regularizing inverse problems via a denoiser. Assume that our goal is to find ${\hat \x}_{MAP}$ by Steepest Descent (SD), and thus our iterative formula should be
\begin{eqnarray}
\label{eq:MAPestH1-SD}
{\hat \x}_{k+1} = {\hat \x}_{k} - \mu \left[ \H^T (\H{\hat \x}_k-\y)  - c\cdot \nabla_{\x}\log p(\x)|_{{\hat \x}_k} \right].
\end{eqnarray}
The term $\nabla_{\x}\log p(\x)$ is known in the statistical literature as the \emph{score function}, being a flow-field that describes the optimal ascent direction over the log of the prior. An old mathematical result, commonly attributed to Miyasawa~\cite{Miyasawa61}, Stein~\cite{stein1981estimation}, or Tweedie~\cite{efron2011tweedie}, and re-exposed in~\cite{kadkhodaie2021stochastic}, proves that
\begin{equation}
    \nabla_{\mathbf{y}} \log p(\mathbf{y}) =
    \frac{D(\mathbf{y}, \sigma_0) - \mathbf{y}}{\sigma_0^2},
\end{equation}
where $\y =\x+\v$ is a noisy version of $\x$ with $\v \sim {\cal N}(\0,\sigma_0 ^2 \I)$, and $D(\mathbf{y}, \sigma_0)$ should be the optimal Minimum Mean Squared Error (MMSE) denoiser, $\mathbb{E}(\mathbf{x} | \mathbf{y})$. A proof of this result is brought in Appendix \ref{App:Score}. 

While it is impossible to obtain the MMSE denoiser (as $p(\mathbf{x})$ is unknown), modern deep learning-based denoisers perform very well (see Figure \ref{fig:PSNrvsTime}), and therefore constitute a good approximation for it. And so, while Equation (\ref{eq:MAPestH1-SD}) expects to use the score function that refers to $p(\x)$, a denoiser can provide an approximation of it that considers a slightly blurry probability density function\footnote{See Appendix \ref{App:Score} for a justification of this claim.} $p(\y)=p(\x)\otimes {\cal N}(\0,\sigma_0 ^2 \I)$. When $\sigma_0$ is small enough\footnote{RED~\cite{RED} suggests to use $\sigma_0 \approx 3-5$ for images with $256\times256$ gray-values.}, this approximation becomes very effective and the resulting algorithm admits the following update rule:
\begin{eqnarray}
\label{eq:MAPestH1-SD-Score}
{\hat \x}_{k+1} = {\hat \x}_{k} - \mu \left[ \H^T (\H{\hat \x}_k-\y)  + c  \left(\x_k - D({\hat \x}_k,\sigma_0)\right)\right],
\end{eqnarray}
which is exactly the SD version of RED~\cite{RED}. 


\subsection{Summary: Denoisers for Solving Inverse Problems}

\begin{figure}[htbp]
    \centering
    \begin{subfigure}{0.3\textwidth}
        \includegraphics[width=\textwidth]{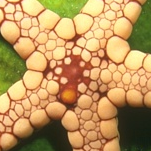}
        \caption{Ground Truth}
    \end{subfigure}
    \begin{subfigure}{0.3\textwidth}
        \includegraphics[width=\textwidth]{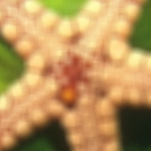}
        \caption{Input $20.83$dB}
    \end{subfigure}
    \begin{subfigure}{0.3\textwidth}
        \includegraphics[width=\textwidth]{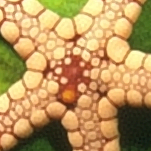}
        \caption{RED (Median) $25.87$dB}
    \end{subfigure}
    \\
    \begin{subfigure}{0.3\textwidth}
        \includegraphics[width=\textwidth]{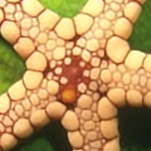}
        \caption{NCSR $28.39$dB}
    \end{subfigure}
    \begin{subfigure}{0.3\textwidth}
        \includegraphics[width=\textwidth]{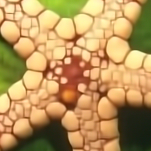}
        \caption{PnP (TNRD) $28.43$dB}
    \end{subfigure}
    \begin{subfigure}{0.3\textwidth}
        \includegraphics[width=\textwidth]{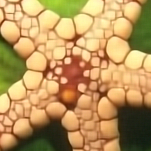}
        \caption{RED (TNRD) $28.82$dB}
    \end{subfigure}
    \caption{Visual comparison of deblurring results by PnP and RED. NCSR~\cite{chen2016trainable} is brought as a reference to compare with. }
    \label{fig:PnP-RED1}
\end{figure}

\begin{figure}[htbp]
    \centering
    \begin{subfigure}{0.3\textwidth}
        \includegraphics[width=\textwidth]{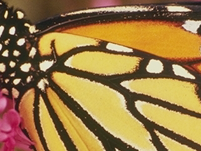}
        \caption{Ground Truth}
    \end{subfigure}
    \begin{subfigure}{0.3\textwidth}
        \includegraphics[width=\textwidth]{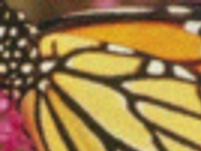}
        \caption{Bicubic $20.68$dB}
    \end{subfigure}
    \begin{subfigure}{0.3\textwidth}
        \includegraphics[width=\textwidth]{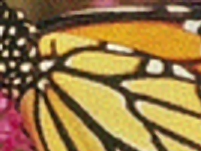}
        \caption{RED (Median) $24.44$dB}
    \end{subfigure}
    \\
    \begin{subfigure}{0.3\textwidth}
        \includegraphics[width=\textwidth]{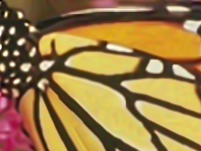}
        \caption{NCSR $26.79$dB}
    \end{subfigure}
    \begin{subfigure}{0.3\textwidth}
        \includegraphics[width=\textwidth]{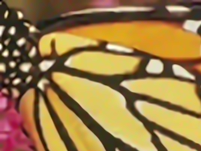}
        \caption{PnP (TNRD) $26.61$dB}
    \end{subfigure}
    \begin{subfigure}{0.3\textwidth}
        \includegraphics[width=\textwidth]{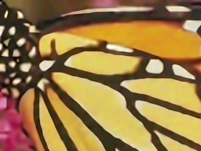}
        \caption{RED (TNRD) $27.39$dB}
    \end{subfigure}
    \caption{Visual comparison of super-resolution (3:1) results by PnP and RED. NCSR~\cite{chen2016trainable} is brought as a reference to compare with.}
    \label{fig:PnP-RED2}
\end{figure}

Figures \ref{fig:PnP-RED1} and \ref{fig:PnP-RED2} present illustrative results of PnP~\cite{PnP}, RED~\cite{RED}, and NCSR~\cite{dong2012nonlocally} for deblurring and single-image super-resolution. Note that while NCSR is specifically tailored to handle these two applications, PnP and RED are unaware of the underlying task, and use a given denoiser. The tests presented employ both a simple median filter and the TNRD denoiser~\cite{chen2016trainable}. Surprisingly, even a plain denoiser as the median filter can provide some recovery effect. More details on these experiments and more results can be found in \cite{RED}. 

PnP and RED have drawn much interest in our community in the past several years. Followup work has been considering a theoretical analysis of the two methods~\cite{chan2016plug,teodoro2017scene,reehorst2018regularization, fletcher2018plug,xu2020provable, sun2021scalable}, deployment of the proposed algorithms in various applications~\cite{sreehari2016plug,brifman2016turning,kamilov2017plug,chen2021deep}, creation of new variants of these two methods~\cite{tirer2018image, teodoro2018convergent,sun2019online, teodoro2019image, sun2019block,hong2019acceleration,cohen2021regularization}, and more. An appealing outlet of this work returns to the unfolding idea discussed in Section \ref{sec:Synergy}: PnP/RED can be used to define well-motivated architectures for solving general inverse problems, by unfolding the proposed algorithms, and then training the repeated denoiser to best serve a series of inverse problems jointly. This way, by plugging in the degradation operator $\H$, a single network can treat a variety of tasks in image processing, built around a core learned denoising engine~\cite{meinhardt2017learning,rick2017one,dong2018denoising,mataev2019deepred,zhang2019deep}. 


\section{Discovery 2: Image Synthesis via Image Denoisers}
\label{sec:Synthesis}

The deep learning revolution has enabled several capabilities that were previously thought to be practically impossible. Among the most intriguing such capabilities is \emph{image synthesis} -- the ability to generate a variety of natural-looking images, without conditioning on any kind of input or initialization.
More formally, the goal of image synthesis is to obtain a random generator 
whose outputs follow the prior distribution of images $\mathbf{x} \sim p(\mathbf{x})$.
Succeeding in this task would testify that we have seized the true distribution of images, and this may aid in solving a variety of  imaging tasks.

A common theme in the definition of such image generators is the need to design of a learned machine $G_\Theta (\z)$, which admits a simply distributed input vector $\z$ (\textit{e.g.}, $\z \sim {\cal N}(0,\I)$) and converts it to a valid sample from $p(\x)$. $G_\Theta (\z)$ is a neural network parameterized by $\Theta$, and various techniques were conceived in the past decade for learning $\Theta$ for best fitting the synthesized results with the destination PDF. In this context, the main tool of interest, which popularized image synthesis, is called GAN -- Generative Adversarial Network~\cite{goodfellow2014generative}. While alternatives to GANs do exist, such as Variational Auto-Encoders (VAE)~\cite{kingma2014autoencoding}, Normalizing Flow (NF) techniques~\cite{rezende2015variational, kingma2018glow}, Autoregressive models~\cite{van2016pixel}, and energy-based methods~\cite{hinton2002training, du2019implicit}, GANs were typically at the lead in image generation. Since their introduction and until recently, GANs have undergone various improvements~\cite{dcgan, pmlr-v70-arjovsky17a, gulrajani2017improved, zhang2019self}, and achieved stellar performance~\cite{brock2018large, karras2020analyzing, sauer2022stylegan}. However, this changed dramatically with the arrival of \emph{diffusion models}~\cite{sohl2015deep, song2019generative, ho2020denoising}. 

GANs, and the other generative models mentioned above, are detached from the topic of image denoising. In contrast, \emph{diffusion models} heavily rely on the \emph{score function} and thus on image denoisers for addressing the task of image synthesis. This recent line of work that started to gain traction, aptly named \emph{score-based generative models}~\cite{song2019generative, song2020improved} or \emph{denoising diffusion probabilistic models}~\cite{sohl2015deep, ho2020denoising}, utilizes deep learning-based denoisers to approximate the score function, which is then used in an iterative algorithm to obtain images $\mathbf{x}$ that are fair samples from the PDF $p(\x)$.

The iterative algorithms used for generation in this context are largely based on Langevin dynamics~\cite{roberts1996exponential,besag2001markov}, a Markov Chain Monte Carlo (MCMC) method with the following transition rule:
\begin{equation}
    \mathbf{x}_{t+1} = \mathbf{x}_t + \alpha \nabla_{\mathbf{x}_t} \log p(\mathbf{x}_t) + \sqrt{2\alpha} \mathbf{z}_t,
\end{equation}
where $\mathbf{z}_t \sim \mathcal{N}(0, \mathbf{I})$, and $\alpha$ is an appropriate small constant.
Initialized randomly, after a sufficiently large number of iterations, and under some mild conditions on $p(\x)$, this process converges to a sampling from the distribution $p(\mathbf{x})$ whose score function is used~\cite{roberts1996exponential}. Intuitively, the algorithm follows the direction of the gradient of the log-probability, climbing from one image to a more probable one. This is a gradient ascent process, and the noise is added in each iteration to provide stochasticity, which effectively leads to sampling from $p(\mathbf{x})$ rather than converging to a local maximum.

While it is tempting to use the true data distribution's score function in Langevin dynamics, a few problems prevent such a use~\cite{song2019generative}. One of the main issues lies with the well-known cardinal manifold assumption~\cite{low_dim_manifold}, which relies on the observation that natural images reside on a low-dimensional manifold in their embedding space. Therefore, for a random initialization of $\mathbf{x}_0$, it holds with probability $1$ that $p(\mathbf{x}_0) = 0$, rendering the score function undefined at best, and without an ability to drift towards the image manifold in subsequent iterations. A possible solution is to approximate $p(\mathbf{x})$ by its slightly blurred counterpart $p(\mathbf{y})$, where $\y=\x+\v$, $\v\sim {\cal N}(0,\sigma^2 \I)$, with a very small $\sigma$~\cite{vincent2011connection}. This resolves the aforementioned problem, as the Gaussian noise distribution has infinite tails. However, in practice, such a Langevin sampling algorithm requires many thousands of iterations to converge~\cite{laumont2021bayesian}, hindering its practical applicability.

\begin{figure}[htbp]
    \centering
    \includegraphics[width=0.9\linewidth]{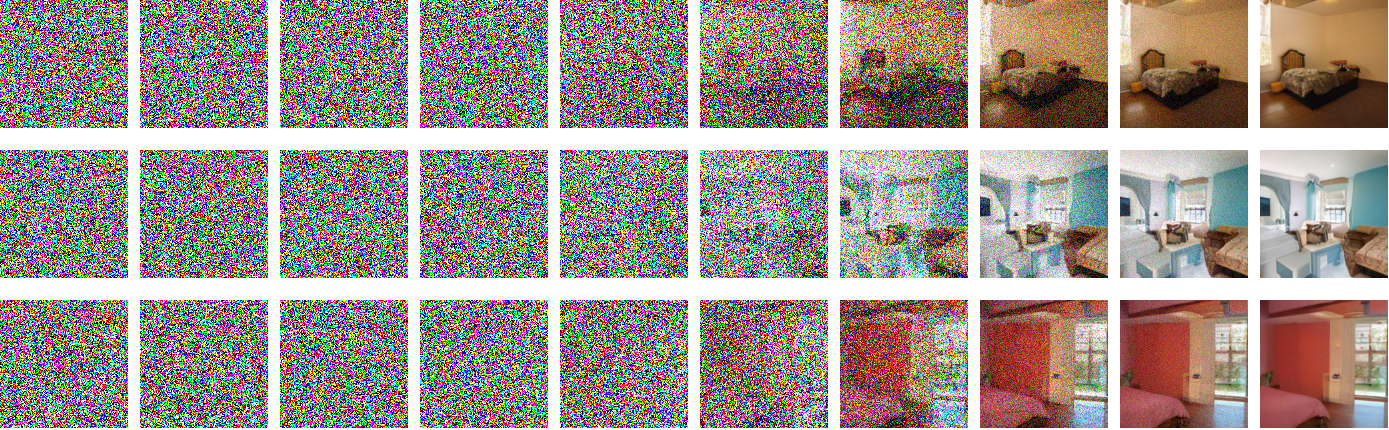}
    \caption{Temporal steps along 3 independent synthesis paths of the Annealed Langevin Dynamics~\cite{song2019generative} algorithm, using a denoiser~\cite{song2020improved} trained on LSUN bedroom~\cite{lsun} images.}
    \label{fig:ALD-Path}
\end{figure}

The authors of~\cite{song2019generative} suggest the \emph{Annealed Langevin Dynamics} (ALD) algorithm\footnote{A very similar algorithm has been proposed in parallel by~\cite{ho2020denoising}. Preceding these two works is the one reported in \cite{sohl2015deep} who proposed a similar process while relying on a different rationale borrowed from statistical physics.}, which considers a sequence of Gaussian noisy image distributions ${p_0(\mathbf{y}), p_1(\mathbf{y}), \dots, p_{L-1}(\mathbf{y}), p_{L}(\mathbf{y})}$ with standard deviations ${\sigma_0 > \sigma_1 > \dots > \sigma_{L-1} > \sigma_L}$. Applying a few iterations of Langevin dynamics for each of the distributions, starting with a very large $\sigma_0$ and ending with a very small $\sigma_L$, enables a faster convergence. Each of these steps is applied using a denoiser that  estimates the score function, and the output of each such process is used to initialize the next. This implies that the synthesis creates a chain of noisy images with diminishing levels of noise, starting with pure canonical Gaussian noise and gradually carving out an image content out of it. Intuitively, this translates to drawing from a wide distribution and then gradually narrowing it, leading to faster sampling and better performance in image generation. Algorithm \ref{alg:ALD} presents this image sampler: The outer loop sweeps through the $L+1$ values of $\sigma$, while the inner loop applies $T$ Langevin steps for each. The score function $\nabla_{\x} \log p_{i}(\x)$, which stands for the $\sigma_i$-blurred PDF of $\x$, is approximated by 
\begin{eqnarray}
\nabla_{\x} \log p_{i}(\x) = \frac{D(\x,\sigma_i) - \x}{\sigma_i^2}.
\end{eqnarray}
Observe that the step size $\alpha$ is modified throughout this process, chosen to be proportional to $\sigma_i^2$. This aligns with the fact that larger $\sigma$ values imply a more regular and smooth PDF, which is easier to sample from.\footnote{A different explanation for this choice of the step size is given in~\cite{song2019generative}, motivated by a desire to better balance the norms of the score versus the additive noise in the Langevin update formula.} Figure \ref{fig:ALD-Path} presents several examples of temporal steps in the ALD process that starts with pure Gaussian noise and ends with a high-quality synthesized image.

\begin{algorithm}[H]
\caption{the Annealed Langevin Dynamics (ALD) algorithm}
\label{alg:ALD}
  \KwIn{$\left\{\sigma_{i}\right\}_{i=0}^{L}$, $\epsilon$, $T$}
  Initialize $\mathbf{x}_0 \sim {\cal N}(0, \I)$

  \For{$i$ $\leftarrow$ $0$ to $L$}{
    $\alpha_i \leftarrow \epsilon \cdot \sigma_i^2/\sigma_L^2$

    \For{$t$ $\leftarrow$ $1$ to $T$}{
     Draw $\z_{t} \sim \mathcal{N}\left(0, \mathbf{I}\right)$ 
     \\
    $\x_t$ $\leftarrow$ $\x_{t-1} + \alpha_i\left[ D(\x_{t-1},\sigma_i) - \x_{t-1}  \right] / \sigma_i^2 + \sqrt{2\alpha_i} \z_t$ 
   }
 $\mathbf{x}_0 \leftarrow \mathbf{x}_T$
 }
 \KwOut{$\mathbf{x}_0$}
\end{algorithm}

\begin{figure}[htbp]
    \centering
    \includegraphics[width=0.8\linewidth]{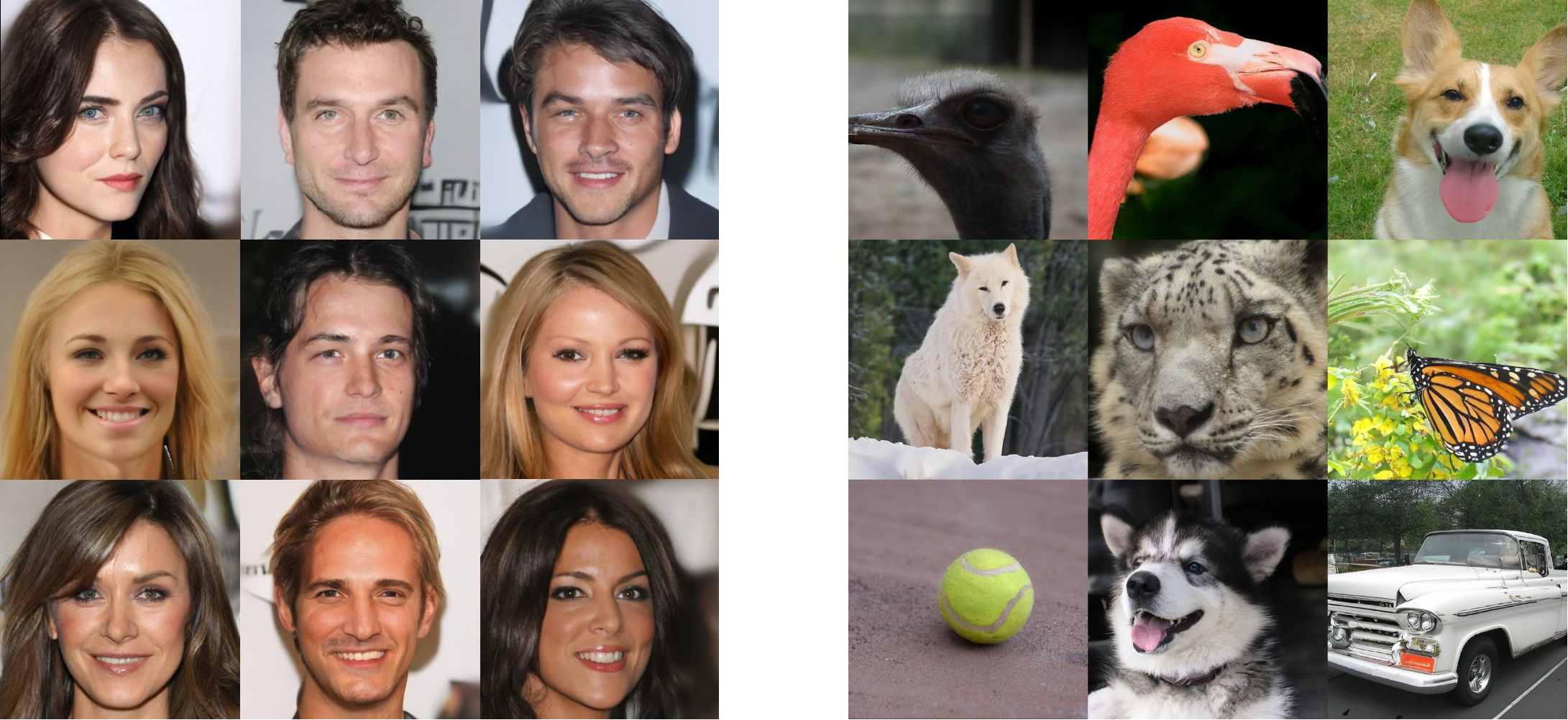}
    \caption{Image generation results for CelebA-HQ~\cite{liu2015celeba} (left) and ImageNet~\cite{imagenet} (right) using score-based denoising diffusion generative models~\cite{song2021sde, guided_diffusion}.}
    \label{fig:diffusion_gen}
\end{figure}

The ALD algorithm sparked a wave of related works~\cite{song2020improved, ho2020denoising, song2021sde, nichol2021improved, vahdat2021score, guided_diffusion, ho2021classifier, kawar2022enhancing, ho2022cascaded} that continually improved the performance of these generative \emph{diffusion models}, eventually surpassing that of GANs~\cite{guided_diffusion}. We show some of their results in Figure \ref{fig:diffusion_gen}.
Nevertheless, these iterative algorithms are still considerably slower than GANs, so substantial work has been invested in improving their speed without compromising significantly on generation quality~\cite{ddim, jolicoeur2021gotta, salimans2022progressive}, often achieving impressive speedup levels.
Diffusion models have since become ubiquitous in many applications~\cite{kawar2022ddrm, nie2022DiffPure, blau2022threat, han2022card, amit2021segdiff, shi2022conditional,sinha2021d2c, kawar2022jpeg}, prompting researchers to prepare surveys of their impact on the image processing field and beyond~\cite{yang2022diffusion, croitoru2022diffusion, cao2022survey}.

\section{ Discovery 3: High Perceptual Quality Image Recovery}
\label{sec:HPC}

We are now stepping into the last and what we believe to be  one of the most exciting topics in the story of image denoisers -- solving general linear inverse problems while striving for perfect perceptual quality, and achieving this with the support of an MMSE denoiser. We start with the simplest inverse problem -- image denoising itself -- and grow from there to more general recovery tasks.


\subsection{Revisiting the Image Denoising Problem}
\label{sec:HPC-Denoising}
We return to the classic image denoising problem, where $\y=\x+\v$ in a given noisy image, $\x\sim p(\x)$ is it's ideal origin, and $\v \sim {\cal N}(0,\sigma_{\y}^2)$ is the AWGN. Our goal is to recover $\x$, but now we change the rules of the game by expecting high perceptual quality results. How could this be achieved? 

Throughout the classical era of denoising, and well into the modern AI days, denoisers were mostly evaluated using the Mean Squared Error (MSE) measure shown in Equation \eqref{eq:MSE} (or tightly related measures such as the Peak Signal-to-Noise Ratio -- PSNR).
As can be seen in Figure~\ref{fig:PSNrvsTime}, MSE has been and still is a commonly used performance measure for denoisers. The MSE metric has several clear benefits: it is zero when the denoiser perfectly recovers the image, it is intuitive to understand, and it produces mathematically elegant results for theoretical analysis, as well as practical considerations such as ease of differentiation for optimization. 

\begin{figure}[htbp]
    \centering
    \includegraphics[width=0.8\textwidth]{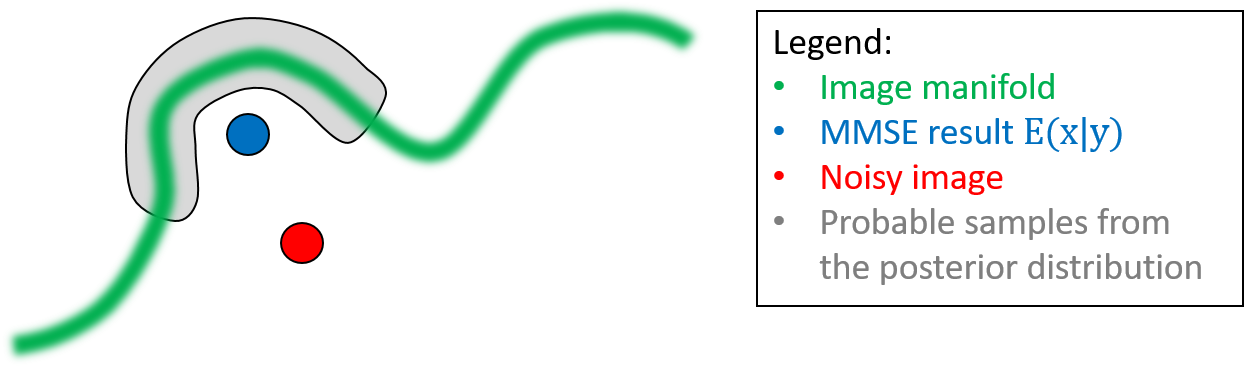}
    \caption{A $2$-dimensional qualitative demonstration of the disadvantages of MMSE denoising. Given a noisy image, the MMSE denoiser falls outside of the image manifold, whereas a posterior sampler would necessarily sample points that reside on it. This leads to better perceptual quality in the denoising results.}
    \label{fig:2d_example}
\end{figure}

However, the MSE distortion measure suffers from a critical shortcoming: As discussed in Section \ref{sec:ProbDef} and in Appendix \ref{App:MMSE}, the best possible result in MSE (MMSE), regardless of the denoising method used to approximate it, would rely on a conditional expectation,
\begin{eqnarray}
{\hat \x}_{MMSE} = \argmin_{{\hat\x}} \mathbb{E}\left( \| \x-{\hat \x} \|_2^2\right) = 
\int_{\x} \x p(\x|\y)d\x = \mathbb{E}\left( \x | \y \right).
\end{eqnarray}
In other words, when optimizing for MSE, our main goal is to get as close as possible to the original image in expectation, and this implies an averaging over all possible solutions, weighted by their posterior probability. 
Thus, depending on the geometry of the image manifold and the severity of the noise, the MMSE solution may tend to be too blurry and of relatively low probability $p({\hat \x}_{MMSE})$, falling outside of the desired manifold. We illustrate this phenomenon in a 2-dimensional example in Figure \ref{fig:2d_example}.

\begin{figure}[htbp]
    \centering
    \includegraphics[width=0.8\textwidth]{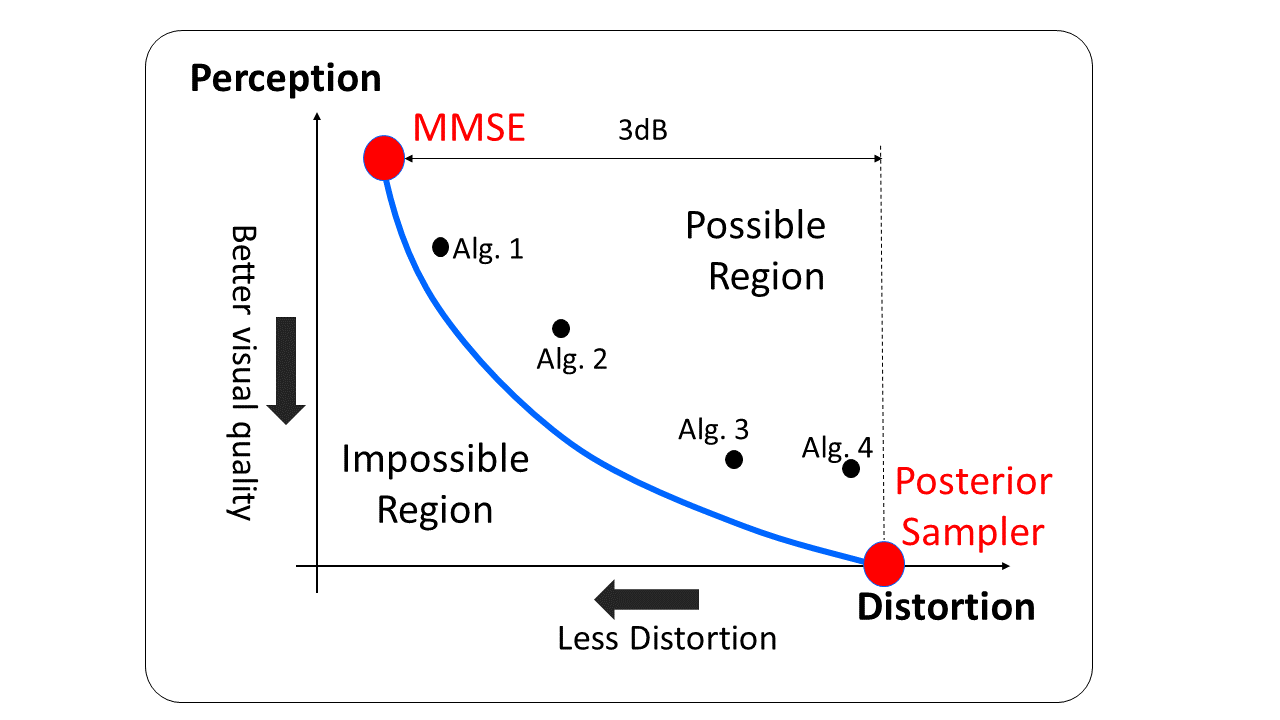}
    \caption{The perception-distortion trade-off~\cite{blau2018PD}: Any recovery algorithm necessarily performs on the blue-curve or above it. On the perception-distortion bound curve, the top-left point refers to the MMSE estimation, while the right-bottom one (or right to it -- see \cite{blau2018PD}) is obtained by a posterior sampler. A gap of $3$dB divides between the two when using the MSE distortion measure.}
    \label{fig:PerceptioDistortionTradeoff}
\end{figure}

Indeed, the fact that MMSE denoising achieves optimal $L_2$ distortion necessarily implies that \emph{perceptual} quality is compromised. The authors of~\cite{blau2018PD} prove the existence of a ``perception-distortion tradeoff'': distortion (of any kind!) and perceptual quality are at odds with each other, and optimizing one necessarily deteriorates the other. In this context, perceptual quality is defined as the proximity between the original image distribution $p(\mathbf{x})$, and the denoised image one $p(\hat{\mathbf{x}})$. Figure~\ref{fig:PerceptioDistortionTradeoff} presents the essence of these findings in~\cite{blau2018PD}. 

With this tension between visual quality and distortion in mind, alternative approaches to MSE were developed over the years, aiming for high perceptual quality denoising~\cite{divakar2017image, dey2020image, ohayon2021high, kawar2021stochastic}. One such technique is to sample from the posterior distribution: given a noisy image $\y$, we aim to develop a denoiser that outputs $\hat{\mathbf{x}} \sim p(\mathbf{x} | \mathbf{y})$, \emph{i.e.}, samples from the posterior distribution of pristine images given the noisy measurement. A successful posterior sampler would achieve perfect perceptual quality, as when marginalizing over $\mathbf{y}$, we get $p(\hat{\mathbf{x}}) = p(\mathbf{x})$.
It is important to notice that this technique involves a subtle paradigm shift -- the denoiser is no longer a deterministic function of the noisy input $\mathbf{y}$, but rather a stochastic one and this implies a multitude of possible solutions.
In the following, we present two pragmatic approaches for approximating posterior sampling behavior.

To traverse the perception-distortion tradeoff, a Waserstein Generative Adversarial Network (WGAN) conditioned on noisy images can be used~\cite{blau2018PD, divakar2017image}. Such a network consists of two main elements: a \emph{generator}, which takes a noisy image as well as a random vector as input, and outputs a denoised image, and a \emph{discriminator}, whose job is to distinguish between denoised and original images. The discriminator is trained to discriminate between the generator's outputs and original images, while the generator optimizes two loss functions: the MSE with respect to the original image, and the ability to ``fool'' the discriminator, thus encouraging its output to ``look like a real image'' in the eyes of the discriminator. These two losses, as proven in~\cite{blau2018PD}, are at odds with one another, and tuning their respective weights in the total loss function translates to the traversal of the perception-distortion tradeoff. This idea is further improved upon by~\cite{ohayon2021high}: instead of requiring low distortion on individual generator samples, the requirement is made on their mean. This results in a loss function that encourages the generator to act as a sampler from the posterior distribution, therefore attaining near-perfect perceptual quality while remaining faithful to the input image.

\begin{figure}[htbp]
    \centering
    \includegraphics[width=0.8\textwidth]{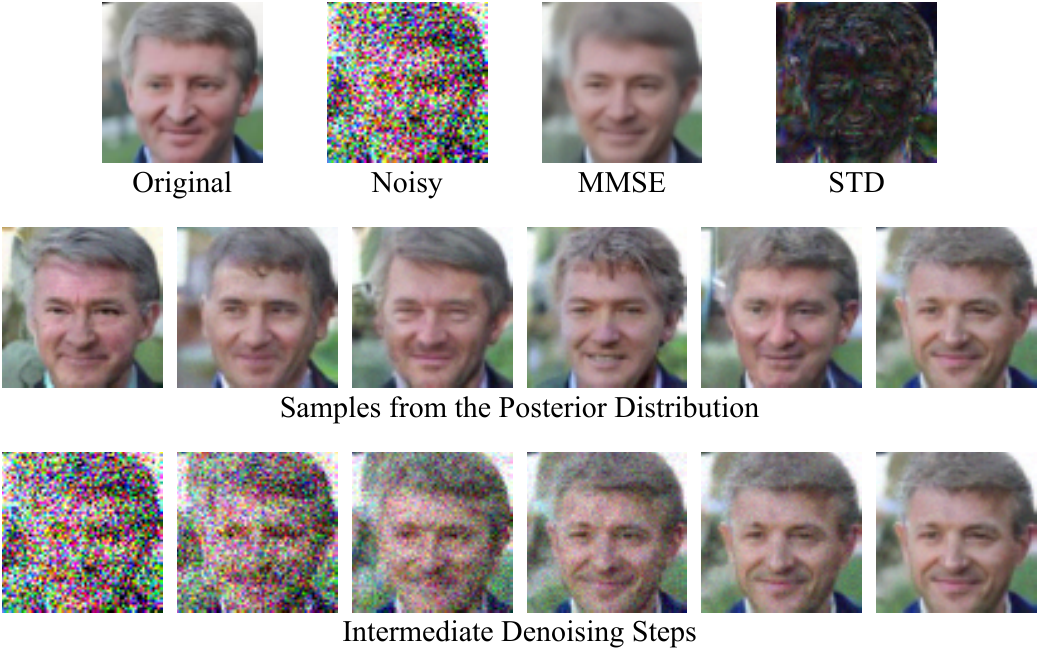}
    \caption{Image denoising using the modified version of Annealed Langevin Dynamics~\cite{kawar2021stochastic}. Top row (left to right): An original image, its noisy version ($\sigma_{\y}=100$), the MMSE-optimized denoiser's result, and the STD of the sampled solutions. 
    Middle row: 6 sampled ALD denoising solutions. Bottom row: 6 intermediate steps within the ALD algorithm.}
    \label{fig:ALD-Denoising}
\end{figure}

An alternative posterior sampling approach, which reconnects with MMSE denoisers, is using the annealed Langevin dynamics algorithm~\cite{song2019generative} presented in the previous section. Recall that ALD uses the score function $\nabla_{{\tilde \x}} \log p_i({\tilde \x})$ to sample from a prior distribution $p_i({\tilde \x})$\footnote{In these notations, $p_i$ stands for a $\sigma_i$-blurred PDF version of the original prior $p(\x)$, and ${\tilde \x}$ is a temporary synthesized image that contains annealing Gaussian noise with variance $\sigma_i^2$.}.  In~\cite{kawar2021stochastic}, the regular ALD algorithm is extended to treat image denoising by analytically  conditioning the score function on a noisy input $\mathbf{y}$ 
-- effectively sampling from the posterior distribution $p_i({\tilde \x} | \mathbf{y})$.
The algorithm is initialized with the noisy input $\mathbf{y}$, which is then gradually denoised using the conditional score function, obtained using the Bayes rule,
\begin{equation}
    \label{eq:score-bayes}
    \nabla_{{\tilde \x}} \log p_i({\tilde \x} | \y) =
    \nabla_{{\tilde \x}} \log \frac{ p(\y|{\tilde \x}) p_i({\tilde \x})} {p(\y)} = 
    \nabla_{{\tilde \x}} \log p_i({\tilde \x}) + 
    \nabla_{{\tilde \x}} \log p(\mathbf{y} | \mathbf{x}_t).
\end{equation}
The term $\nabla_{{\tilde \x}} \log p_i({\tilde \x})$ is the regular score function which can be approximated by an MSE-trained denoiser.
As for the other term, $\nabla_{{\tilde \x}} \log p(\y | {\tilde \x})$, observe that this likelihood can be rewritten by exploiting two facts: (i)  $\y = \x+\v$ is the noisy image ($\v\sim {\cal N}(0,\sigma_{\y}^2\I$), and (ii) ${\tilde \x} = \x+{\z}$ is the annealed solution (${\z} \sim {\cal N}(0,\sigma_i^2 \I)$), and thus
\begin{eqnarray}
\label{eq:LikihoodCond}
p(\y | {\tilde \x}) & = & p(\y - {\tilde \x}| {\tilde \x}) 
\\ \nonumber
& = & p(\x+\v - \x - \z| {\tilde \x})
\\ \nonumber
& = & p(\v - \z| {\tilde \x}).
\end{eqnarray}
If we assume statistical independence between the measurements' noise and the annealing one, $\v - \z$ becomes a plain Gaussian vector. However, its conditioning on the knowledge of ${\tilde \x}$ leads to a dead-end, since this image contains $\z$ in it. The alternative, as developed in \cite{kawar2021stochastic}, is to construct the annealing noise such that $\v - \z$ is statistically independent of both $\z$ and ${\tilde \x}$. This can be obtained by breaking the measurements' noise $\v$ into small fragments, and assume that their partial accumulations constitute the annealing noise in each of the stages. Thus, $\v - \z$ is a white Gaussian noise that has no correlation with the noise $\z$, nor with the target image $\x$. Put in other words, this likelihood expression becomes simple when considering $\mathbf{y}$ to be an even more noisy version of ${\tilde \x}$. This in turn makes $p(\mathbf{y} | {\tilde \x})$ a simple white Gaussian distribution of the form ${\cal N}(0,(\sigma_{\y}^2 - \sigma_i^2)\I)$. Therefore,
\begin{equation}
    \label{eq:score-bayes2}
    \nabla_{{\tilde \x}} \log p_i({\tilde \x} | \mathbf{y}) =
    \nabla_{{\tilde \x}} \log p_i({\tilde \x}) + 
    \frac{\mathbf{y} - {\tilde \x}}{\sigma_\mathbf{y}^2 - \sigma_i^2}.
\end{equation}
Plugging this modification into ALD turns Algorithm \ref{alg:ALD} into an image denoiser. Beyond its ability to attain near-perfect perceptual quality, this approach has the advantage of not requiring any special model training. Crucially, this finding shows that simple MSE denoiser training is more powerful than originally thought -- not only can it approximate MMSE denoiser behavior, but it can also perform denoising by posterior sampling under the Langevin dynamics scheme. Figure \ref{fig:ALD-Denoising} presents a denoising result by the above-described method. Several observations are in order from this figure: 
\begin{itemize}
\item The generated results are indeed of very high perceptual quality; 
\item Running ALD several times results with different solutions, all valid and yet diverse -- see the STD image that exposes the uncertainty within the task being solved; 
\item Denoising $\y$ directly by $D(\y,\sigma_{\y})$ leads to better MMSE but poorer perceptual quality; 
\item The figure also shows the evolving solution within the ALD steps, and as can be seen, the noise in $\y$ is effectively peeled layer by layer. 
\end{itemize}

\subsection{High Perceptual Quality Solution to Inverse Problems}
\label{sec:HPC-IP}

We now expand our discussion by returning to general linear inverse problems of the form $\y = \H \x+\v$, where $\H \in \mathbb{R}^{M\times N}$ is a known matrix, $\v \in \mathbb{R}^M$ is AWGN, and $\y\in \mathbb{R}^M$ is the given measurement vector. Our goal is to propose novel solutions to these problems while striving for high perceptual quality. 

The above discussion on the perception-distortion tradeoff is not limited to image denoising, but also applies to more general inverse problems~\cite{blau2018PD}. There too, potential solvers need to tradeoff distortion metrics (\textit{e.g.} MSE) versus perception measures (\textit{e.g.} the distribution shift between real images and the obtained solutions). Indeed, MSE in these cases may become far more challenging as an optimization goal due to the ill-posedness of the inverse problems. Consider, as an example, an inpainting problem in which the bottom half of the image is given and the goal is to recover the top part. The MMSE solution in this case necessarily averages all possible completions, resulting in a very blurry outcome. More broadly, optimizing for MSE in this context would result in a clear regression-to-the-mean, which is significantly more pronounced in under-determined inverse problems than in image denoising.

Successful inverse problem solvers, such as the Plug-and-Play Prior~\cite{PnP} and RED~\cite{RED} algorithms mentioned in Section \ref{sec:IP},  aim for a Maximum-a-Posteriori (MAP) solution to the inverse problem at hand, rather than MMSE. While these methods achieve impressive results, 
the MAP solution can be improved upon in terms of perceptual quality without compromising on distortion performance~\cite{blau2018PD}. This is due to the deterministic nature of MAP solvers -- a solver that aims for best perceptual quality should necessarily be stochastic in order to account for the multiple possible solutions to the given problem~\cite{ohayon2022reasons}.

\begin{figure}[htbp]
    \centering
    \includegraphics[width=0.8\textwidth]{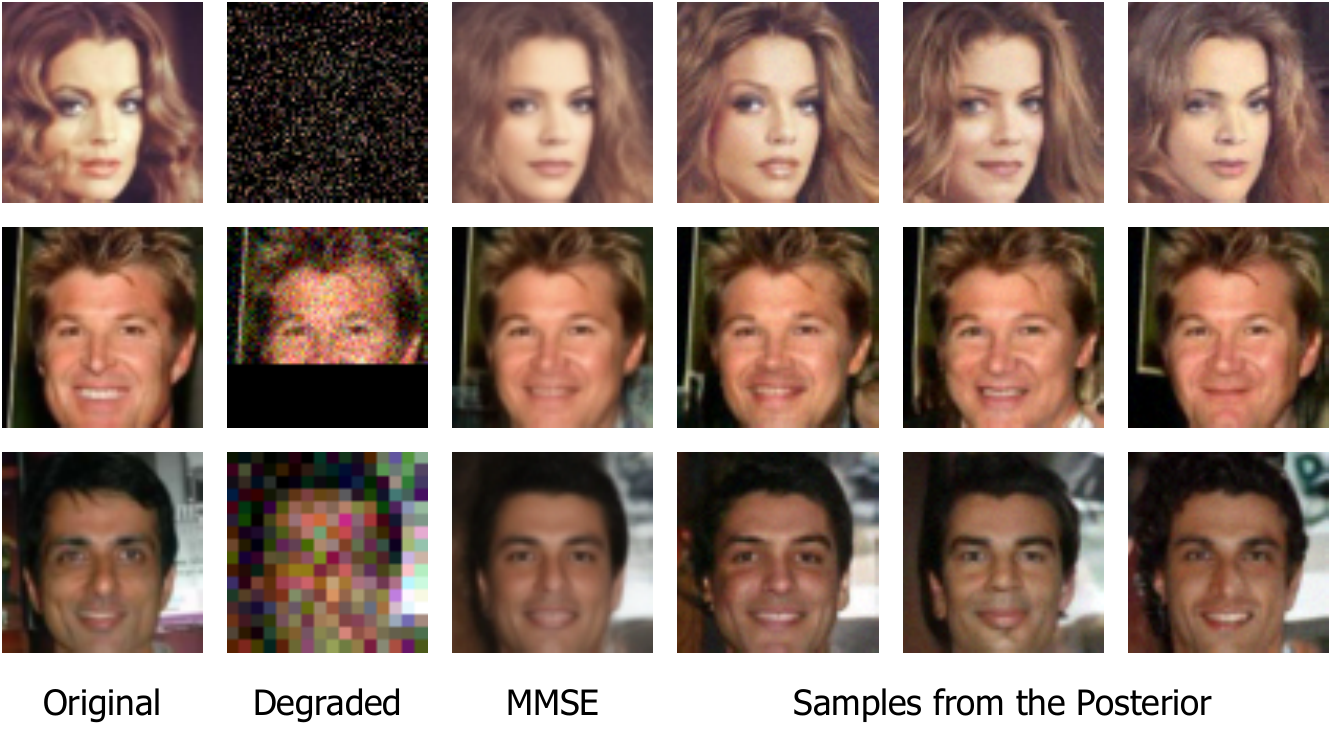}
    \caption{Comparison of an MMSE result with samples from the posterior distribution using SNIPS~\cite{kawar2021snips}. Note the subtle improvements in perceptual quality from MMSE to the posterior samples, especially in the finer details such as the hair. The comparison is conducted on $64\times64$ pixel images from CelebA~\cite{liu2015celeba}, on the problems of compressive sensing, inpainting, and $4\times$ super-resolution (top-to-bottom).}
    \label{fig:mse_vs_posterior}
\end{figure}

Similar to the image denoising case, stochastically sampling from the posterior distribution achieves perfect perceptual quality in general inverse problems. Following the road paved in the previous section, an appealing way to approximate such sampling would be to follow Equation (\ref{eq:score-bayes}), using a generative diffusion model and augmenting the score by an analytical term that conditions on the observed measurement $\mathbf{y}$. This idea has been initially suggested by~\cite{song2021sde,kadkhodaie2021stochastic} for handling noiseless linear inverse problems, and later extended to the more general case in~\cite{kawar2021snips,kawar2022ddrm, chung2022improving, chung2022diffusion, meng2022diffusion}. Below we describe the essence of the proposed approach in SNIPS~\cite{kawar2021snips}. Visual examples of this method in action are brought in Figure \ref{fig:mse_vs_posterior} for several inverse problems. 

Our goal is to obtain a closed-form expression for the term $\nabla_{{\tilde \x}} \log p(\y | {\tilde \x})$ in Equation (\ref{eq:score-bayes}). We use the following two relationships: (i)  $\y = \H \x+\v$ is the noisy measurement ($\v\sim {\cal N}\left(0,\sigma_{\y}^2\I\right)$), and (ii) ${\tilde \x} = \x+\z$ is the annealed solution ($\z \sim {\cal N}(0,\sigma_i^2 \I)$). The likelihood function can be simplified to
\begin{eqnarray}
\label{eq:LikihoodCondH1}
p(\y | {\tilde \x}) & = & p(\y - \H{\tilde \x}| {\tilde \x}) 
\\ \nonumber
& = & p(\H\x+\v - \H\x - \H\z| {\tilde \x})
\\ \nonumber
& = & p(\v - \H\z| \x + \z).
\end{eqnarray}
As in the denoising case in Equation (\ref{eq:LikihoodCond}), statistical independence between $\v$ and $\z$ cannot be assumed due to the dependency on ${\tilde \x}$. The alternative, as shown by SNIPS~\cite{kawar2021snips} relies again on a delicate connection between these two random entities, obtained by a decoupling of the  measurements' equation via an Singular Value Decomposition (SVD) of the degradation matrix $\mathbf{H}= \U \boldsymbol{\Sigma} \V^T$: 
\begin{eqnarray}
\label{eq:LikihoodCondH2}
p(\y | {\tilde \x}) & = &  
p(\v - \H\z| \x + \z) 
\\ \nonumber & = & 
p(\U^T \v - \boldsymbol{\Sigma}\V^T \z| \V^T\x + \V^T\z)
\\ \nonumber & = & 
p({\hat \v} - \boldsymbol{\Sigma}{\hat \z}| {\hat \x} + {\hat \z}) 
\\ \nonumber & = &
\prod_k p({\hat v}_k - s_k {\hat z}_k | {\hat x}_k + {\hat z}_k).
\end{eqnarray}
The second row in the above equation is obtained by transforming the term $\v - \H\z$ by the matrix $\U^T$, and similarly transforming $\x + \z$ via a multiplication with $\V^T$.
As these are unitary matrices, the transformations applied do not change the statistics.
Considering the transformed vectors $\U^T\y = {\hat \y}$, $\V^T\x = {\hat \x}$, $\V^T\z = {\hat \z}$ and $\U^T\v = {\hat \v}$ leads to the third row in the above equation. This joint probability can be decoupled into a separable Gaussian distribution if we choose each entry ${\hat v}_k - s_k {\hat z}_k$ to be independent of ${\hat z}_k$, just as practiced in the denoising case, and this time while taking into account the singular value $s_k$.
This algorithm, fully described in~\cite{kawar2021snips},  demonstrates considerable success in a number of inverse problems (see Figure \ref{fig:mse_vs_posterior}), and already has several followup works~\cite{kawar2022ddrm, chung2022improving, chung2022diffusion, meng2022diffusion}.


We should mention that an alternative to all the above exists, in which one simply adds the corrupted measurements $\mathbf{y}$ as an input to the denoising model itself, effectively conditioning the entire generative process on $\mathbf{y}$~\cite{saharia2022image, saharia2022palette, whang2022deblurring}.
This approach requires designing and training a separate denoiser for each inverse problem, as the denoiser would need to implicitly learn the connection between the images and their corresponding measurements for the specific problem at hand. Interestingly, this approach requires pairs of images, $\mathbf{x}$ and $\mathbf{y}$, in its training, but does not utilize knowledge of the degradation model itself (\textit{e.g.}, the matrix $\mathbf{H}$).
This property allows this alternative approach to generalize beyond clearly formulated inverse problems, and handle tasks such as stylization, JPEG-deblocking, and more.

\begin{figure}[htbp]
    \centering
    \includegraphics[width=\linewidth]{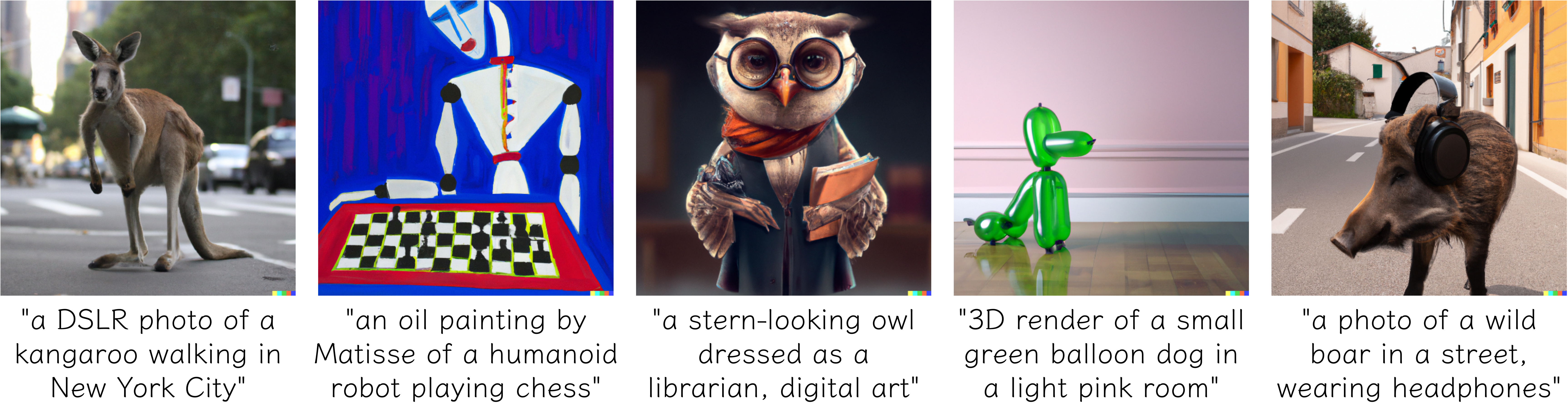}
    \caption{Examples of synthesized images using DALL-E 2~\cite{ramesh2022hierarchical}, a text-to-image generative denoising diffusion model. The input conditioning text is written below each image.}
    \label{fig:text2img}
\end{figure}

A particularly interesting case is when $\mathbf{y}$ is a textual description of the image contents. By conditioning the denoiser model on such text, the generative diffusion process allows users to perform \emph{text-to-image generation}~\cite{ramesh2022hierarchical, saharia2022photorealistic, rombach2022high, balaji2022ediffi}.
This unprecedented capability became instantly popular, as users were able to synthesize high-quality images by simply describing the desired result in natural language, as we demonstrate in Figure \ref{fig:text2img}.
These models have become a centerpiece in an ongoing and quickly advancing research area, as they have been adapted for image editing~\cite{kawar2022imagic, mokady2022null}, object recontextualization~\cite{ruiz2022dreambooth, gal2022image}, 3D object generation~\cite{poole2022dreamfusion}, and more~\cite{ho2022imagenvideo, jain2022vectorfusion, pan2022extreme, zhu2022exploring}.


\section{Conclusion}
\label{sec:conclusion}

\begin{figure}[htbp]
    \centering
    \includegraphics[width=1\linewidth]{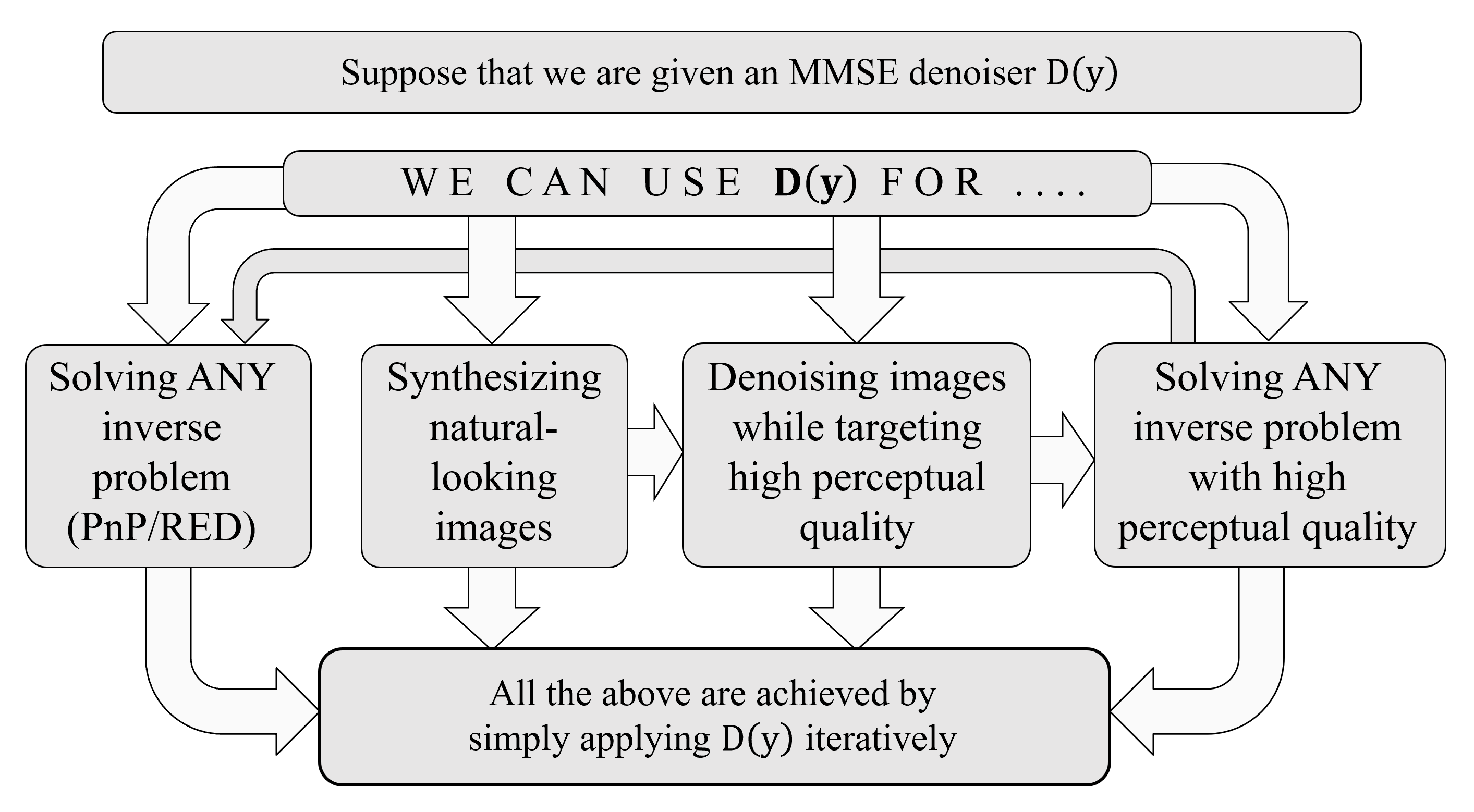}
    \caption{A summary of the main message of this paper: an MMSE denoiser is key in synthesizing images and solving inverse problems. Interestingly, there is a great unexplored proximity between PnP and RED algorithms~\cite{PnP,RED} and the more recent, diffusion-based, techniques for getting high perceptual quality solutions for inverse problems~\cite{kawar2021snips, chung2022diffusion, meng2022diffusion}.}
    \label{fig:Summary-Image}
\end{figure}

Removal of white additive Gaussian noise from an image is a fascinating topic, both because it poses a very interesting engineering challenge, and even more so, because it creates new opportunities in image processing and machine learning. In this paper we highlight these two branches of activities. The first half of the paper concentrates on the design of such denoisers, with a particular interest on the impact of the AI revolution on this field. The second half of the paper features the usefulness of such image denoisers for handling other tasks, such as image synthesis and solving inverse problems while targeting high-perceptual quality solutions. Figure \ref{fig:Summary-Image} encapsulates this part of the story in a block diagram. 

Much remains to be done in this domain, in better understanding how to design appropriate MMSE denoisers, and in harnessing them to other tasks beyond the ones described in this paper, such as compression, segmentation, and more. 
More broadly, there are so many opportunities and challenges in better understanding, designing, and proposing creative usage of image denoisers. 

\newpage
\appendix

\section{Derivation of the MMSE Estimation}
\label{App:MMSE}

Consider an ideal image $\x$ drawn from the probability density function $p(\x)$, and assume that we are given a measurement of it, $\y$, related to it via the conditional probability $p(\y|\x)$. Our goal is to find the estimator ${\hat \x}=f(\y)$ that minimizes the expected mean-squared-error, 
\begin{eqnarray}
\label{eqApp:MSE}
MSE = \mathbb{E}\left( \| \x-{\hat \x} \|_2^2~ |~\y \right)=\mathbb{E}\left( \| \x- f(\y) \|_2^2~|~\y\right) = \int \| \x- f(\y) \|_2^2 p(\x|\y)d\x.
\end{eqnarray}
Observe that this expectation is taken with respect to the unknown image $\x$, while considering $\y$ as known. In order to minimize the above measure, we take a derivative of this expression with respect to $f(\y)$ and null it,  
\begin{eqnarray}
\label{eqApp:MSEderivative}
\frac{d}{df(\y)} \int \left\| \x- f(\y) \right\|_2^2 p(\x|\y)d\x =  - 2 \int \left(\x- f(\y) \right) p(\x|\y)d\x =0.
\end{eqnarray}
This results in
\begin{eqnarray}
\label{eqApp:MSEderivative2}
\int \x p(\x|\y)d\x = \int  f(\y) p(\x|\y)d\x =  f(\y) \int  p(\x|\y)d\x = f(\y).
\end{eqnarray}
The last step on the right-hand-side relies on the fact that $\int  p(\x|\y)d\x=1$. Thus, we get the familiar closed-form solution for the MMSE estimation~\cite{MMSE}, 
\begin{eqnarray}
\label{eq:MMSE-final}
f_{MMSE}(\y) = \int_\x \x p(\x|\y) d\x = \mathbb{E}\left( \x | \y \right).
\end{eqnarray}
As a final step, as the posterior is not directly accessible, we may use the Bayes rule~\cite{bayes} and write 
\begin{eqnarray}
\label{eq:MMSE-final-2}
f_{MMSE}(\y) = \int_\x \x \frac{p(\y|\x) p(\x)}{p(\y)} d\x = \int_\x \x \frac{p(\y|\x) p(\x)}{\int_\x p(\y|\x) p(\x) d\x} d\x,
\end{eqnarray}
where this formula uses the ingredients we started with -- $p(\y|\x)$ and $p(\x)$.


\section{A Closer Look at the Evolution of Priors}
\label{App:Priors}

Using the Gibbs distribution form, $p(\x)=c\cdot \exp\{-\rho(\x)\}$, we shift our focus from the probability density function $p(\x)$ to it's corresponding energy function $\rho(\x)$. Table \ref{tab:Priors} brings a list of possible analytical expressions for $\rho(\x)$ as evolved in the image processing literature. Below we describe each of these options briefly, adopting the context of solving a general linear inverse problem of the form $\y=\H\x+\v$ with the assumptions that $\v \sim {\cal N}(\0,\sigma ^2 \I)$ and $\H$ is a full-rank known matrix of size $m \times N$ ($m<N$). The MAP estimation in this case is given by 
\begin{eqnarray}
\label{eq:MAP-IP}
{\hat \x}_{MAP} & = & \argmin_{\x} \left[ \frac{\|\H \x-\y\|_2^2}{2\sigma^2}  - \log\left( p(\x) \right) \right] \\ \nonumber & = & \argmin_{\x} \left[ \|\H\x-\y\|_2^2  + c \cdot \rho(\x) \right].
\end{eqnarray}
Notice that $2\sigma^2$ was absorbed into the constant: $c = 2\sigma^2$. Armed with this expression, let's consider each of the choices in Table \ref{tab:Priors} and explore its implications. Before diving into these options, observe that without the regularization provided by $\rho(\x)$, the above optimization becomes an ill-posed Least-Squares problem with infinitely many possible solutions. Thus, the added prior serves as an important regularization, pushing towards a single (and hopefully, meaningful) solution. 

\begin{itemize}

\item {\bf Energy regularization:} If $\H^T \H$ cannot be inverted, the most obvious algebraic remedy would be to add a constant to its diagonal, resulting with the regularized solution  ${\hat \x}_{MAP} = (\H^T \H +c\I)^{-1} \H^T \y$. This is exactly the solution offered by the choice $\rho(\x) = \|\x\|_2^2$, and when the constant $c$ is taken to $0$, this leads to the familiar pseudo-inverse solution ${\hat \x}_{MAP} = \H^\dagger \y$. While mathematically appealing, this option does not yield satisfactory visual results~\cite{schafer1981constrained}.   

\item {\bf Spatial Smoothness:} It is well-known that adjacent pixels in natural images are more likely to be of smoothly varying values. Thus, penalizing a deviation from such a smoothness property seems well-justified~\cite{banham1997digital,lagendijk2009basic}. Plugging the option $\rho(\x) =\|\L \x\|_2^2$ into the MAP expression leads to the closed-form solution ${\hat \x}_{MAP} = (\H^T \H +c\L^T\L)^{-1} \H^T \y$, which is very-closely related to the well-known Wiener filter~\cite{wiener}.

\item {\bf Optimally Learned Transform:} Given a large enough dataset of images, we could fit a multivariate Gaussian ${\cal N}(\0,\R)$ to them by adjusting the second moment. The assumed zero mean is easily obtained by subtracting the mean image from the given data. PCA~\cite{pearson1901pca} or Karhunen-Lo\'eve Transform (KLT)~\cite{loeve,karhunen,castleman1996digital,jolliffe2016principal} offer a clear computational path towards this moment matrix $\R$ as the auto-correlation matrix of the available data. When the expression $\rho(\x) = \x^T \R^{-1} \x $ is plugged into the MAP estimation, we come back to the Wiener filter, this time as ${\hat \x}_{MAP} = (\H^T \H +c\R^{-1})^{-1} \H^T \y$. Note that the same treatment could emerge from this formulation -- $\rho(\x) = \|\T \x\|_2^2 = \x^T \R^{-1} \x $, where $\T$ is the corresponding transform that should be applied on $\x$, and clearly $\T^T\T = \R^{-1}$. 

\item {\bf Weighted Smoothness:} All the above options suffer from the same difficulty -- they produce overly smoothed results. In retrospect, the reason is obvious: non-smooth behavior is heavily penalized and thus not encouraged, which results with smeared edges. A way to overcome this difficulty is to produce a weight map that describes the local smoothness tendency -- regions in which smoothness is believed to be correct should be assigned with a high weight, while low weight should be given to regions suspected to be textured or edges~\cite{saint1991adaptive,chu1998edge}. By constructing a diagonal matrix $\W$ that contains the above weights as the main diagonal, and using the choice \& $\rho(\x) = \|\L \x\|_{\W}^2$, the MAP estimation becomes ${\hat \x}_{MAP} = (\H^T \H +c\L^T\W \L)^{-1} \H^T \y$. This is a spatially adaptive solution, dependent on the local weights. One may consider an iterative approach where the temporary solution ${\hat \x}_{MAP}$ is leveraged to update the weights and then ${\hat \x}_{MAP}$ is re-computed. This interesting option leads to the robust statistics alternative discussed next~\cite{black1998robust}. 

\end{itemize}

\noindent Before proceeding with the other prior options, we would like to draw the readers' attention to the fact that all the above choices correspond to the core assumption that the probability density function $p(\x)$ is a multivariate Gaussian. The obtained visual results of these techniques expose the fact that this Gaussianity assumption is not adequate, and indeed, later research in image processing turned to non-Gaussian and heavy-tailed alternative distributions, which we discuss next. 

\begin{itemize}

\item {\bf Robust statistics:} Here is a simple experiment -- take any natural image, apply a Laplacian on it, and gather a histogram of the resulting values. This histogram is likely to look as a heavy-tailed probability density function of a form similar to $c\cdot \exp(-|x|^\alpha )$ with $\alpha\ll 2$. This is exactly the deviation from Gaussianity referred to above. Thus, the robust statistics alternative~\cite{huber1981robust,geman1984stochastic,geman1992constrained,charbonnier1997deterministic,rousseeuw2011robust} suggests a replacement of the $L_2$-norm of $\L\x$ by $L_1$ or, more broadly, by functions of the form $\1^T \mu\{\L \x\}$ (\textit{e.g.} $\mu(x)=|x|^\alpha$). Notice that from here on, closed-form MAP solution cannot be obtained, and iterative minimization strategies are necessary. 

Adopting a different point of view, robust statistics considers pixel on edges and textures regions as outliers to the Gaussian distribution, and thus use robust estimation techniques for their better handling.  

\item {\bf Total-Variation (TV):} The same motivation as described above led to this brilliant PDE formulation of spatial smoothness, $\rho(\x) = \int_{v\in\Omega} |\nabla \x(v)| dv$, which accumulates the length of the spatial gradients instead of their squares~\cite{rudin1992nonlinear}. In its discretized form, its treatment is very similar to the robust-statistics option. However, TV has very different roots, providing a geometrically oriented edge-preserving measure of smoothness -- see various extensions of this line of work in~\cite{beck2009fast,gilboa2009nonlocal,chambolle2011first,aggarwal2016hyperspectral}. 

\item {\bf Other PDE-based options:} While TV applies an $L_1$-norm on the spatial gradients, more general options can be envisioned, in which the accumulation is spatially adaptive, orientation sensitive, geometrically faithful, and more~\cite{perona1990scale,catte1992image,sochen1998general,weickert1998pde,Guishard20022pde}. Starting with the seminal anisotropic diffusion method by Perona and Malik~\cite{perona1990scale}, various such methods of the form $\rho(\x) = \int_{v\in\Omega} g\left[\nabla \x(v),\nabla^2 \x(v)\right] dv$ were proposed and perfected over the years, forming an exciting sub-field of mathematically oriented image processing that relies on the vast knowledge in partial differential equations. 

\item {\bf Field-of-Experts (FoE):} Let us return to the robust statistics option described above and enrich it by considering a mixture of such distributions, $\rho(\x) = \sum_k \lambda_k \1^T  \mu_k \{\L_k \x\} $. This implies the need to define a series of functions $\mu_k$ and their corresponding weights $\lambda_k$. FoE suggests to learn these elements from an image dataset, thus better fitting the assumed prior to natural images. While earlier work on FoE~\cite{Roth2005FieldsOE} suggested a patch-based maximum-likelihood learning approach, later efforts~\cite{chen2016trainable} brought a deep-learning alternative tools to this fitting. 

\item {\bf Wavelet sparsity:} The idea of relying on transform coefficients for constructing $\rho(\x)$ has already been explored in the context of the KLT. The emergence of the Wavelet transform in the late 80's brought a new way of thinking about signals and images, offering an elegant and more effective multi-scale representation that relies on non-linear approximation~\cite{donoho1994ideal,donoho1995noising,coifman1995translation,ghael1997improved,mallat1999wavelet,luisier2010fast,chang2000adaptive,portilla2003image,luisier2007new,zhang2008wavelets,goossens2009removal}. Wavelets offer a concise description of the data with as few as possible coefficients, this way giving birth to the central notion of sparsity. This translates well to the proposed prior $\rho(\x) = \|\W \x\|_1$ that promotes fewer non-zero dominant Wavelet coefficients. 

As an interesting side note, if we are handling the image denoising problem -- \textit{i.e.} $\H=\I$ in Equation (\ref{eq:MAP-IP}) -- and the Wavelet transform matrix $\W$ is unitary, the solution ${\hat \x}_{MAP}$ has a closed-form solution, obtained via a soft-shrinkage~\cite{donoho1994ideal,donoho1995noising}. 

\item {\bf Self-similarity:} So far we described two primary forces that promote simplicity in image content -- spatial smoothness and representation sparsity. Self-similarity is a third such force that has been recognized as central by series of contributions, starting with the seminal Non-Local-Means (NLM) algorithm~\cite{buades2005non}, and heavily relied upon by the famous BM3D~\cite{dabov2007image} and other algorithms~\cite{mairal2009non,tasdizen2009principal,manjon2010adaptive,chatterjee2011patch,wang2013adaptive,mosseri2013combining,talebi2013global, salmon2014poisson,liu2018non}. Self-similarity stands for the assumption that any given (small-enough) patch in an image is likely to find very similar ones in the image support, and thus treating these together somehow is likely to lead to better recovery. More specifically, the expression we bring here as an illustration, 
\begin{eqnarray}
\rho(\x) = \sum_k \sum_{j\in \Omega(k)}d\{\R_k \x, \R_j \x\},
\end{eqnarray}
sweeps through the image support, extract a patch in location $k$ by the operator $\R_k \x$, and finds all its corresponding matches $j\in \Omega(k)$. Forcing proximity between $\R_k \x$ and the patches $\R_j \x$ induces a strong regularization over the unknown image $\x$.

\item {\bf Sparsity methods:} While the notion of sparsity has already been exploited by wavelets, later work took this idea and strengthened it by considering redundant and learned representations. Under the assumption that ideal images can be described as linear combinations of atoms from a pre-specified dictionary $\D$, \textit{i.e.}, $\x=\D\alpha$, forcing sparsity on the representation via the term $\|\alpha\|_0$ provides an appealing and computationally feasible choice for $\rho(\x)$~\cite{bruckstein2009sparse,elad2010sparse}. Vast work along these lines has been done, considering global dictionaries and later local (patch-based) ones, leading to various very successful recovery algorithms~\cite{elad2006image,Elad2005SimultaneousCA,aharon2006ksvd,mairal2007multiscale,Mairal2008SparseRF,mairal2009non,dupe2009proximal,yu2011solving,dong2011sparsity,dong2012nonlocally,giryes2014sparsity,dong2015image,egiazarian2015single}. 

\item {\bf Low-Rank assumption:} The last member to enter the Pantheon of image priors for image processing relies on a low-rank assumption over groups of similar patches. This idea is closely related to the self-similarity force described above, and in fact builds on top of it. Given a set of closely related patches, instead of forcing proximity between them, one may gather these as columns in a matrix and force a low-rank structure, implying that all these patches are spanned by few main directions. Several very strong recovery algorithms leveraged this idea in various forms, while exploiting theoretical analysis that ties the low-rank requirement to the nuclear-norm~\cite{wright2009robust,candes2011robust}. By summing these norms over such groups, $ \rho(\x) = \sum_k \|\X_{\Omega(k)}\|_* $, a very potent regularization is obtained~\cite{WNNM_2014,yair2018multi}.  
\end{itemize}

\setlength{\tabcolsep}{5pt}
\begin{table*}
\begin{center}
\caption{Evolution of priors for images.}
\label{tab:PriorsAgain}
\begin{tabular}{|l|l|l|l|}
\hline 
Years & Core concept & Formulae for $\rho(\cdot)$ & Representative \\
  &  & & Reference \\
\hline
$\sim$ 1970 & Energy regularization & $\|\x\|_2^2$ & \cite{schafer1981constrained}\\
\hline
1975-1985 & Spatial smoothness & $ \|\L \x\|_2^2$ or $ \|\D_v \x\|_2^2 + \|\D_h \x\|_2^2$ & \cite{lagendijk2009basic} \\
\hline
1980-1985 & Optimally Learned Transform & $\|\T \x\|_2^2 = \x^T \R^{-1} \x $ & \cite{castleman1996digital} \\
& & where $\T / \R$ is learned via PCA &  \\
\hline
1980-1990 & Weighted smoothness & $\|\L \x\|_{\W}^2$ & \cite{saint1991adaptive} \\
\hline
1990-2000 & Robust statistics & $\1^T \mu\{\L \x\} $ & \cite{black1998robust} \\
& & \textit{e.g.}, Hubber-Markov & \\
\hline
1992-2005 & Total-Variation & $\int_{v\in\Omega} |\nabla \x(v)| dv$ & \cite{rudin1992nonlinear}  \\
& & or $ \1^T \sqrt{|\D_v \x|^2 + |\D_h \x|^2}$ &  \\
\hline
1987-2005 & Other PDE-based options & $\int_{v\in\Omega} g\left[\nabla \x(v),\nabla^2 \x(v)\right] dv$ & \cite{weickert1998pde} \\
\hline
2005-2009 & Field-of-Experts & $\sum_k \lambda_k \1^T  \mu_k \{\L_k \x\} $ & \cite{Roth2009FieldsOE} \\
\hline
1993-2005 & Wavelet sparsity & $\|\W \x\|_1$ & \cite{donoho1995noising} \\
\hline
2000-2010 & Self-similarity & 
$\sum_k \sum_{j\in \Omega(k)}d\{\R_k \x,\R_j \x\}$ & \cite{buades2005non,dabov2007image} \\ 
\hline
2002-2012 & Sparsity methods & $\|\alpha\|_0 ~s.t.~ \x=\D\alpha$ & \cite{bruckstein2009sparse} \\
\hline
2010-2017 & Low-Rank assumption & $ \sum_k \|\X_{\Omega(k)}\|_* $ & \cite{WNNM_2014} \\ \hline
\end{tabular}
\end{center}
\end{table*}

\noindent As a summary, the above-described evolution of the priors has served as the skeleton of image processing, forming the consistent progress of this field over the years. This evolution is characterized by four major and interconnected trends:
\begin{enumerate}
    \item A migration from the familiar Gaussian distribution to the less intuitive heavy-tailed ones; 
    \item A departure from $L_2$ to sparsity-promoting norms, such as the $L_1$;
    \item A drift from linear approximation techniques (\textit{e.g.} PCA) to non-linear ones (\textit{e.g.} wavelets and sparse modeling); and above all, 
    \item A replacement of axiomatic expressions with learned ones.
\end{enumerate}


\section{Landmark Denoisers over the Years}
\label{App:DenoisingAlgorithms}

In Figure \ref{fig:PSNrvsTime} we brought a graph showing the PSNR performance of landmark denoising algorithms over the years. Below we provide more information on these techniques for completeness of this study. For each of these we bring the full reference, describe the core algorithmic idea, and provide the PSNR denoising performance on the BSD68 dataset ($\sigma=25$). We should note that in choosing the methods to include in this list we restricted the scope to ones that report of BSD68 results. 

\begin{itemize}

\item {\bf KSVD~\cite{elad2006image} [28.28dB]:} Elad, M., \& Aharon, M. (2006). Image denoising via sparse and redundant representations over learned dictionaries. IEEE Transactions on Image processing, 15(12), 3736-3745.

This method decomposes the noisy image into fully overlapping patches, and denoises each by sparse approximation via OMP~\cite{pati1993orthogonal}, while learning an over-complete dictionary. The denoised image is obtained by returning the cleaned patches to their original locations while averaging them over the overlaps and with a weighted version of the noisy image. 

\item {\bf BM3D~\cite{dabov2007image} [28.57dB]:} Dabov, K., Foi, A., Katkovnik, V., \& Egiazarian, K. (2007). Image denoising by sparse 3-D transform-domain collaborative filtering. IEEE Transactions on Image Processing, 16(8), 2080-2095.

This algorithm extracts all fully overlapping patches from the noisy image and gathers similar patches into 3D blocks. Denoising is performed by transforming these blocks, forcing sparsity, and then transforming the sparse outcome back to the image domain. The denoised image is obtained by returning the patches to their original locations while averaging over the overlaps. This process is ran twice, where the first round serves for an initial cleaning that improves the patch correspondences for the later round. 

\item {\bf FoE~\cite{Roth2009FieldsOE} [27.77dB]:} Roth, S., \& Black, M. J. (2009). Fields of experts. International Journal of Computer Vision, 82(2), 205-229.

FoE (appeared originally in 2005~\cite{Roth2005FieldsOE}) builds a generic prior that mixes several regularizers (called "experts"). The prior's parameters are learned via a contrastive divergence penalty and MCMC sampling. The image denoising itself is obtained by an iterative algorithm that computes the MAP estimation.

\item {\bf LSSC~\cite{mairal2009non} [28.70dB]:} Mairal, J., Bach, F., Ponce, J., Sapiro, G., \& Zisserman, A. (2009). Non-local sparse models for image restoration. CVPR (pp. 2272-2279).

This algorithm combines the sparse representations (as in KSVD) and non-local similarity (as in BM3D) concepts. It decomposes the noisy image into fully overlapping patches and groups similar patches together. These groups of patches are denoised by a joint sparse approximation that forces the same support over a learned dictionary. The denoised image is obtained by returning the patches to their original locations and averaging over the overlaps. 

\item {\bf EPLL~\cite{zoran2011learning} [28.71]:} Zoran, D., \& Weiss, Y. (2011). From learning models of natural image patches to whole image restoration. ICCV (pp. 479-486).

EPLL models the distribution of image patches as a Gaussian Mixture Model (GMM), and learns its parameters off-line with a dataset of clean images. Denoising with EPLL is a MAP estimation, posed as a minimization problem with a regularizer that consists of a sum of patch log-likelihoods. This task is solved by applying quadratic half-splitting and iterating over patch denoising and the whole image accumulation steps. 

\item {\bf MPL~\cite{burger2012image} [28.96dB]:} Burger, H. C., Schuler, C. J., \& Harmeling, S. (2012, June). Image denoising: Can plain neural networks compete with BM3D?. CVPR (pp. 2392-2399). 

This is the first effective deep-learning based method for image denoising. This method extracts all fully overlapped patches as in classical algorithms, and filters each patch by applying a multi-layer Perceptron (fully connected network). The reconstructed image is obtained by returning the patches to their locations and averaging over the overlapping regions. 

\item {\bf CSF~\cite{Schmidt2014ShrinkageFF} [28.74dB]: } 
Schmidt, U., \& Roth, S. (2014). Shrinkage fields for effective image restoration. CVPR (pp. 2774-2781).

This algorithm poses a MAP estimation problem using a product of cascaded shrinkage functions as a local prior. The parameters of these functions are learned from a dataset as in FoE. The algorithm solves the obtained optimization by half-quadratic splitting 
and iterating between local and global optimization steps. 

\item {\bf WNNM~\cite{WNNM_2014} [28.83dB]:} 
Gu, S., Zhang, L., Zuo, W., \& Feng, X. (2014). Weighted nuclear norm minimization with application to image denoising. CVPR (pp. 2862-2869).

This method decomposes an incoming image into fully overlapping patches and groups similar patches arranging them as columns of a matrix. Denoising of the patches is performed by forcing the rank of the constructed matrices to be small by minimizing the matrix nuclear norm. The reconstructed image is obtained  by returning the patches to their original locations while averaging the overlaps. 

\item {\bf TNRD~\cite{chen2016trainable} [28.92dB]:} Chen, Y., \& Pock, T. (2016). Trainable nonlinear reaction diffusion: A flexible framework for fast and effective image restoration. IEEE Transactions on Pattern Analysis and Machine Intelligence, 39(6), 1256-1272.

This method builds on the FoE method, by unfolding the minimization over its prior and this way defining a parametric trainable network. Once the architecture is defined, TNRD trains this neural network end-to-end in a supervised fashion using clean/noisy pairs of images. Denoising is a simple inference of the resulting machine.

\item {\bf DnCNN~\cite{DnCNN} [29.23dB]:} 
Zhang, K., Zuo, W., Chen, Y., Meng, D., \& Zhang, L. (2017). Beyond a gaussian denoiser: Residual learning of deep cnn for image denoising. IEEE Transactions on Image Processing, 26(7), 3142-3155.

This is the first deep learning method that outperforms classical algorithms by a considerable gap. It filters images by applying a convolutional neural network. The network architecture is composed of convolutional layers followed by batch normalizations and ReLU. The network is trained end-to-end using a dataset consisting of noisy/clean image pairs.

\item {\bf IRCNN~\cite{zhang2017learning} [29.15dB]} Zhang, K., Zuo, W., Gu, S., \& Zhang, L. (2017). Learning deep CNN denoiser prior for image restoration. CVPR (pp. 3929-3938).

This method is similar to DnCNN, but uses dilated convolutions within the architecture in order to enlarge the receptive field, thus creating an opportunity for a non-local processing. The network is trained end-to-end using a dataset consisting of noisy/clean image pairs.

\item {\bf NLRN~\cite{liu2018non} [29.41dB]:} Liu, D., Wen, B., Fan, Y., Loy, C. C., \& Huang, T. S. (2018). Non-local recurrent network for image restoration. NeurIPS.

This method incorporates the non-local similarity concept into a convolutional recurrent neural network in an explicit way. The denoising is done by recurrently applying convolutions and weighted averaging of similar regions (as in NLM~\cite{buades2005non}) in the feature space. The network is trained end-to-end using a dataset consisting of noisy/clean image pairs.

\item {\bf MVCNN~\cite{Liu2018MultilevelWF} [29.41dB]:} Liu, P., Zhang, H., Zhang, K., Lin, L., \& Zuo, W. (2018). Multi-level wavelet-CNN for image restoration. CVPR Workshop (pp. 773-782).

This algorithm incorporates the wavelet sparsity concept into the deep learning approach by combining the U-Net architecture with the multi-level wavelet transform. It replaces the downsampling and upsampling U-Net layers with the 2$D$ discrete wavelet transform and it's inverse. The network is trained end-to-end using a dataset consisting of noisy/clean image pairs.

\item {\bf N3Net~\cite{Pltz2018NeuralNN} [29.30dB]:} Plötz, T., \& Roth, S. (2018). Neural nearest neighbors networks. NeurIPS.

This method combines the deep learning approach with the non-local self-similarity concept. This method introduces a differentiable continuous relaxation of the $k$-nearest neighbor (KNN) selection rule and uses it as a building block within the neural network. N3Net's architecture interleaves convolutional blocks with KNN relaxation blocks. The convolutional blocks perform denoising, while the KNN parts augment the feature maps by breaking them into patches, applying patch matching, and finding $k$-nearest neighbors for each patch. The network is trained end-to-end using a dataset consisting of noisy/clean image pairs.

\item {\bf FFDNet~\cite{zhang2018ffdnet} [29.19dB]:} Zhang, K., Zuo, W., \& Zhang, L. (2018). FFDNet: Toward a fast and flexible solution for CNN-based image denoising. IEEE Transactions on Image Processing, 27(9), 4608-4622.

While the architecture of this deep learning method resembles DnCNN, it enlarges the receptive field by reshaping the incoming image into four downsampled sub-images that are simultaneously fed into the network. The network is trained end-to-end using a dataset consisting of noisy/clean image pairs.  

\item {\bf FOCNet~\cite{Jia2019FOCNetAF} [29.38dB]} Jia, X., Liu, S., Feng, X., \& Zhang, L. (2019). Focnet: A fractional optimal control network for image denoising. CVPR (pp. 6054-6063).

This algorithm suggests a novel architecture to replace the one used by DnCNN, relying on an interpretation of residual neural networks as solvers of dynamical systems. While DnCNN refers to integer-order ordinary differential equation, FOCNet's architecture poses a fractional optimal control (FOC) problem that translates into better connectivity. The algorithm for solving the equation is implemented using a feed-forward convolutional neural network whose parameters are learned using a dataset of images.

\item {\bf RIDNet~\cite{anwar2019real} [29.34dB]:} Anwar, S., \& Barnes, N. (2019). Real image denoising with feature attention. CVPR (pp. 3155-3164).

This algorithm introduces attention modules to a neural network whose architecture includes convolutional layers and skip connections. This attention is designed to capture feature dependencies and enhance the weight of important correspondences. The network is trained end-to-end using a dataset of clean/noisy image pairs.

\item {\bf GCDN~\cite{Valsesia2019DeepGI} [29.35dB]:} Valsesia, D., Fracastoro, G., \& Magli, E. (2020). Deep graph-convolutional image denoising. IEEE Transactions on Image Processing, 29, 8226-8237.

This method combines the deep-learning approach with graph modeling. The GCDN architecture includes convolutional and graph-convolutional layers. While regular convolutional layers catch local interrelations between pixels, the graph-convolution ones are designed to capture the non-local dependencies. Each graph-convolutional layer dynamically applies non-local aggregation (graph-convolution). The graph is constructed via a $k$-nearest neighbor whose vertices are feature vectors. Each vertex is connected to the $k$ most similar ones in terms of the $L_2$ norm. The network is trained using a dataset of images. 

\item {\bf SwinIR~\cite{liang2021swinir} [29.50dB]:} Liang, J., Cao, J., Sun, G., Zhang, K., Van Gool, L., \& Timofte, R. (2021). Swinir: Image restoration using swin transformer. CVPR (pp. 1833-1844).

This algorithm incorporates non-locality into convolutional deep learning architecture using shifted window (Swin) transformer modules~\cite{Liu2021SwinTH}. These modules are designed to compute local self-attention in shifted windows, this way exploitig non-local self-similarity. The SwinIR architecture is trained end-to-end using a dataset consisting of noisy/clean image pairs.  

\item {\bf DRUNet~\cite{Zhang2020PlugandPlayIR} [29.48dB]:} Zhang, K., Li, Y., Zuo, W., Zhang, L., Van Gool, L., \& Timofte, R. (2021). Plug-and-play image restoration with deep denoiser prior. IEEE Transactions on Pattern Analysis and Machine Intelligence.

This denoiser is a bias-free~\cite{mohan2019robust} neural network that combines ResNet~\cite{resnet} and U-Net~\cite{unet}. Its architecture includes convolutions, downscaling and upscaling layers, and skip connections. The network is trained using a dataset of images. 

\end{itemize}


\section{Approximation of the \emph{Score Function} by an MMSE Denoiser}
\label{App:Score}

In Section \ref{sec:IP} we brought the definition of the \emph{score function}, $\nabla_{\x}\log p(\x)$, and its approximation via a denoiser. Here we bring the derivation of this result, following the work by Miyasawa~\cite{Miyasawa61}, Stein~\cite{stein1981estimation}, and  Tweedie~\cite{efron2011tweedie}. 

Consider an ideal image $\x\in \mathbb{R}^N$ drawn from the Probability Density Function (PDF) $p(\x)$. Assume that $\y$ is a noisy version of it, $\y =\x+\v$, where $\v \sim {\cal N}(\0,\sigma_0 ^2 \I)$. The PDF of $\y$ can be obtained by a marginalization, 
\begin{eqnarray}
p(\y) = \int_{\x} p(\y|\x)p(\x)d\x = \left[\frac{1}{2\pi\sigma_0^2}\right]^{N/2}\int_{\x} \exp\left\{\frac{-1}{2\sigma_0^2}  \|\y-\x\|_2^2\right\}p(\x)d\x.
\end{eqnarray}
In the above we used the fact that $p(\y|\x) \sim {\cal N}(\x,\sigma_0 ^2 \I)$. The obtained relationship expresses $p(\y)$ as a convolution between the original prior $p(\x)$ and an isotropic zero-mean Gaussian of width $\sigma_0$. Taking a derivative of both sides with respect to $\y$ results in the following: 
\begin{eqnarray}
\nabla_{\y} p(\y) & = & \left[\frac{1}{2\pi\sigma_0^2}\right]^{N/2}\int_{\x} \nabla_{\y}\exp\left\{\frac{-1}{2\sigma_0^2}  \|\y-\x\|_2^2\right\}p(\x)d\x 
\\ \nonumber& = & 
\frac{1}{\sigma_0^2} \cdot \left[\frac{1}{2\pi\sigma_0^2}\right]^{N/2}\int_{\x} (\y - \x)\exp\left\{\frac{-1}{2\sigma_0^2}  \|\y-\x\|_2^2\right\}p(\x)d\x
\\ \nonumber& = & 
\frac{1}{\sigma_0^2} \int_{\x} (\y - \x)p(\y|\x)p(\x)d\x.
\end{eqnarray}
Dividing both sides by $p(\y)$ leads to
\begin{eqnarray}
\frac{\nabla_{\y} p(\y)}{p(\y)} = \nabla_{\y} \log p(\y) & = & 
\frac{1}{\sigma_0^2} \int_{\x} (\x - \y)\frac{p(\y|\x)p(\x)}{p(\y)}d\x 
\\ \nonumber & = &  
\frac{1}{\sigma_0^2} \int_{\x} (\x - \y)p(\x|\y)d\x .
\end{eqnarray}
Opening and rearranging the above expression leads to our final result: 
\begin{equation}
\nabla_{\y} \log p(\y) = \frac{1}{\sigma_0^2} \left[ \int_{\x} \x p(\x|\y)d\x - \y \int_{\x}  p(\x|\y)d\x\right] =    \frac{1}{\sigma_0^2}\left[D(\mathbf{y}, \sigma_0) - \mathbf{y}\right],
\end{equation}
where $D(\mathbf{y}, \sigma_0)$ should be the optimal Minimum Mean Squared Error (MMSE) denoiser, $\mathbb{E}(\mathbf{x} | \mathbf{y})$. Thus, access to an approximation of the score function $\nabla_{\x}\log p(\x)$ can be obtained by using a small value $\sigma_0$, and evaluating the above expression with a given denoiser.  


\newpage
\bibliographystyle{siamplain}
\bibliography{refs_new}

\end{document}